\documentclass[12pt,preprint]{aastex}
\usepackage{longtable}
\usepackage{color}
\usepackage[section] {placeins}
\everymath{\rm}
\everydisplay{\sf}
\usepackage{graphicx}

\begin{document}

\title{The curious conundrum regarding sulfur abundances in planetary nebulae }

\author{R.B.C. Henry}
\affil{H.L. Dodge Department of Physics \& Astronomy, University of Oklahoma, Norman, OK 73019, USA}
\email{henry@nhn.ou.edu}

\author{Angela Speck}
\affil{Department of Physics \& Astronomy, University of Missouri, Columbia, MO 65211 USA}
\email{speckan@missouri.edu}

\author{Amanda I. Karakas}
\affil{Research School of Astronomy \& Astrophysics, Mount Stromlo Observatory, ACT 2611, Australia}
\email{akarakas@mso.anu.edu.au}

\author{Gary J. Ferland}
\affil{Department of Physics \& Astronomy, University of Kentucky, Lexington, KY 40506 USA}
\email{gjferland@gmail.com}

\and

\author{Mason Maguire}
\affil{H.L. Dodge Department of Physics \& Astronomy, University of Oklahoma, Norman, OK 73019, USA}
\email{mr.magu@ou.edu}

\begin{abstract}

Sulfur abundances derived from optical emission line measurements and ionization correction factors in planetary nebulae are systematically lower than expected for the objects' metallicities. We have carefully considered a large range of explanations for this ``sulfur anomaly", including: (1) correlations between the size of the sulfur deficit and numerous nebular and central star properties; (2) ionization correction factors which under-correct for unobserved ions; (3) effects of dielectronic recombination on the sulfur ionization balance; (4) sequestering of S into dust and/or molecules;  and (5) excessive destruction of S or production of O by AGB stars. It appears that all but the second scenario can be ruled out. However, we find evidence that the sulfur deficit  is generally reduced but not eliminated when S$^{+3}$ abundances determined directly from IR measurements are used in place of the customary sulfur ionization correction factor. We tentatively conclude that the sulfur anomaly is caused by the inability of commonly used ICFs to properly correct for populations of ionization stages higher than S$^{+2}$.

\end{abstract}


\section{Introduction}\label{intro}

The sulfur anomaly was first identified by \citet[][hereafter HKB04]{hkb04} and refers to the systematically lower sulfur abundances exhibited by Galactic planetary nebulae (PNe) compared to H~II regions of the same metallicity, where metallicity is gauged by oxygen abundance. Specifically, HKB04 computed S abundances by observing S$^+$ and S$^{+2}$ abundances directly and then applying a sulfur ionization correction factor (ICF) to correct for the contribution of unobserved sulfur ions to the total abundance. HKB04 were unable to establish an exact cause but speculated that the problem is likely brought about by the use of a sulfur ICF which fails to correct adequately for unobserved ions. 

The sulfur anomaly has also been recognized and discussed more recently by \citet{bs06}, \citet{pbs06}, \citet{shaw10}, \citet{kh11}, \citet{green11}, and \citet{rodriguez11}, but still no clear consensus on its cause has developed. \citet{pbs06} and \citet{bs06} used ISO measurements in the IR to obtain S$^{+3}$ ion abundances of 26 PNe. They then obtained the total sulfur abundance by computing the sum of the S$^+$, S$^{+2}$, and S$^{+3}$ abundances for each object, using abundance values found in the literature for the first two ions. Their resulting S abundances still fell below expected levels, and they suggested that the missing gas-phase sulfur might be sequestered into dust. (We address this idea below in Section~\ref{speck}.) Similarly,  \citet{shaw10} used both deep ground-based optical spectra along with Spitzer IR measurements to determine sulfur abundances in 14 PNe in the Small Magellanic Cloud. Like \citet{pbs06} and \citet{bs06} they found that the direct observation of S$^{+3}$ did not solve the problem. They also claimed that the sulfur abundance problem worsened for their three most highly ionized objects. This appears to be direct evidence suggesting that ionization stages above S$^{+3}$ are significantly populated yet unaccounted for by a standard sulfur ICF. \citet{rodriguez11} computed models of PNe with various physical conditions and then subjected the output spectral line strengths to the same analysis as real observed measurements of PNe. They found that the S abundance inferred from the predicted line strengths can exceed the input abundance by 0.05~dex or fall below it by 0.3~dex and stressed that caution should be used when using an ICF to determine S abundances. Finally, \citet{green11} and \citet{kh11} discussed the sulfur anomaly but offer no solutions to the problem.

\begin{figure}[h]
   \includegraphics[width=6in,angle=270]{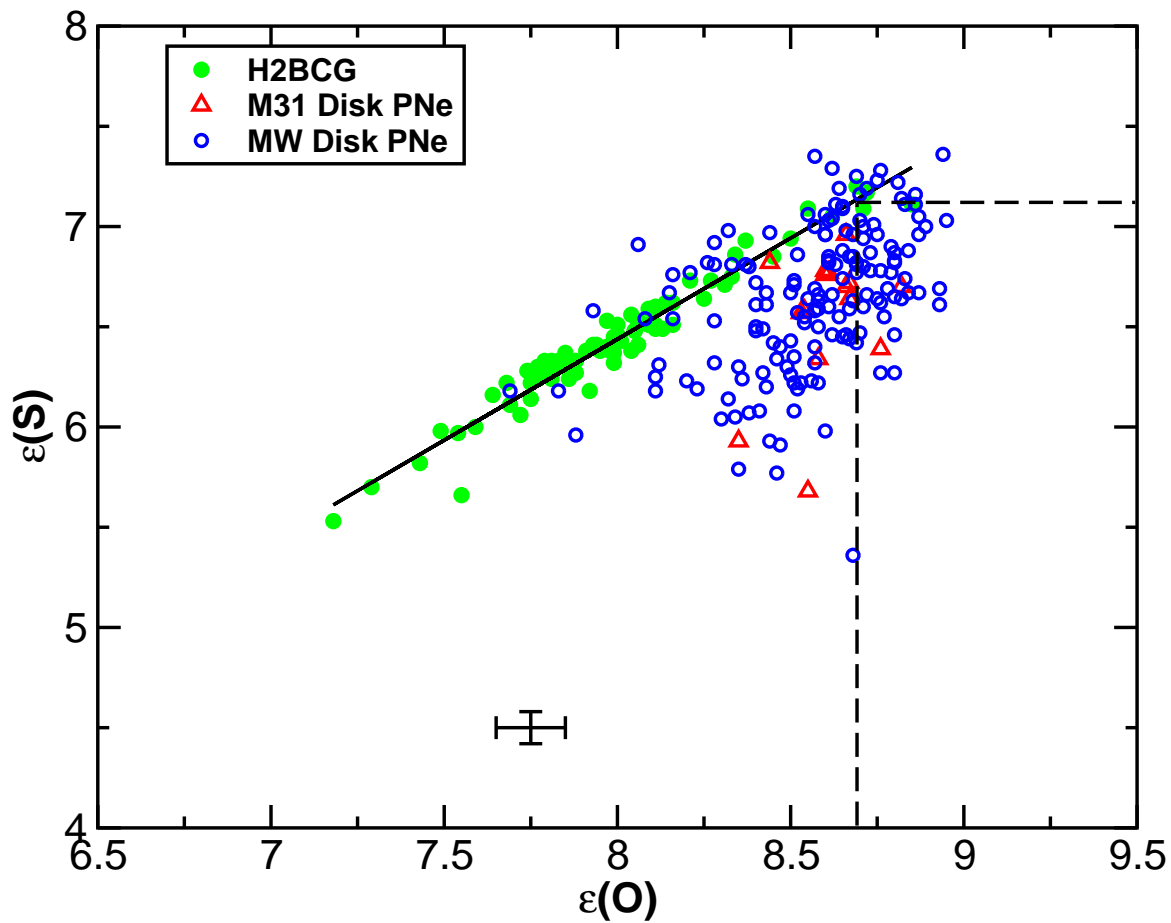}
   \caption{Plot of $\epsilon(S)$ versus $\epsilon(O)$ for a large sample of MWG and M31 disk planetary nebulae (open symbols) along with a sample of H~II regions and blue compact galaxies selected from the literature (H2BCG; filled symbols). The solid line is a least squares fit to the H2BCG sample.   The dashed lines show solar abundances from \citet{asplund09}. Average uncertainties for the S and O abundances of the PN sample, determined by propagating line measurement uncertainties through the abundance-determining step, are shown.}
\label{svo_initial}
\end{figure}

\begin{figure}[h]
   \includegraphics[width=6in,angle=270]{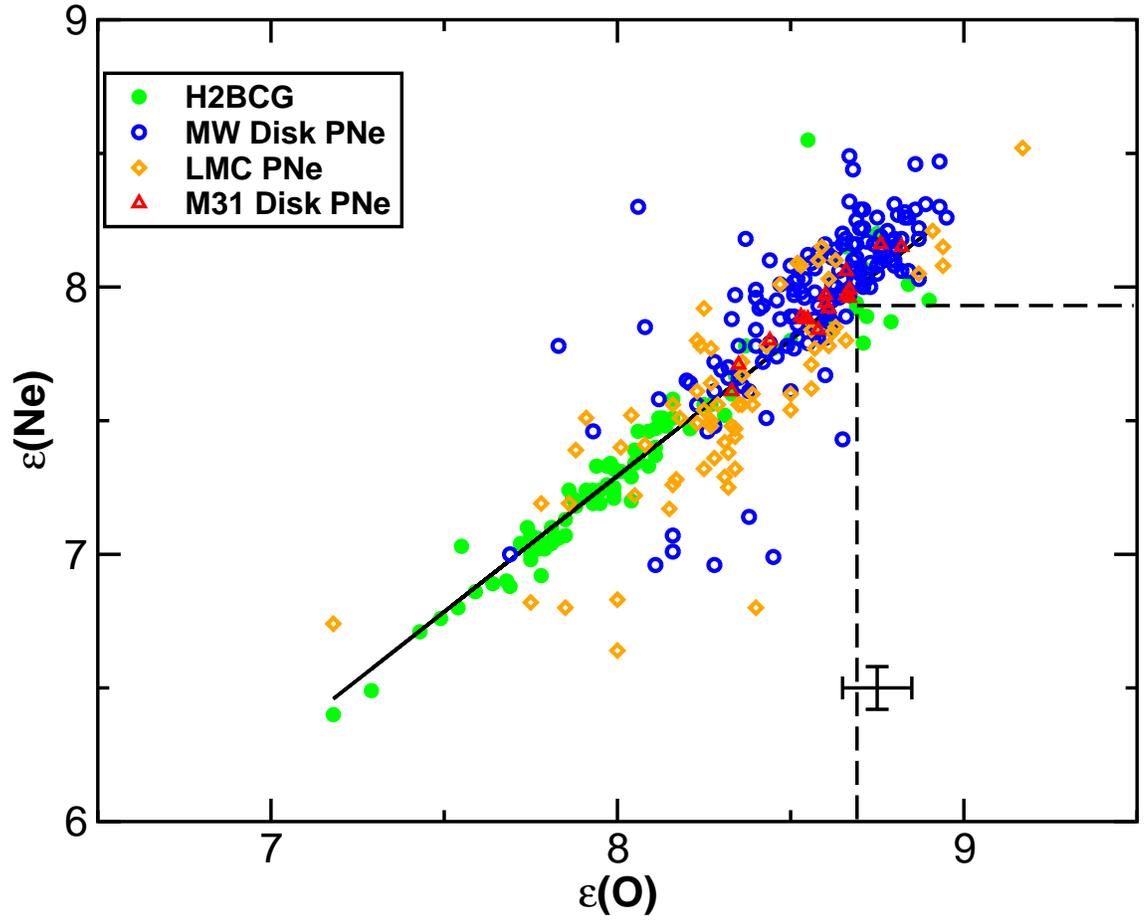}
   \caption{Same as Fig.~\ref{svo_initial} but for $\epsilon(Ne)$ versus $\epsilon(O)$. We have also added PNe from the LMC.}
\label{nevo_initial}
\end{figure}

The sulfur anomaly can be seen clearly in Figure \ref{svo_initial}, where we have plotted $\epsilon(S)$\footnote{$\epsilon(X)=12+log(X/H)$} versus $\epsilon(O)$ for a large sample of PNe in the disks of the Milky Way and M31 galaxies (open symbols) as well as a combined sample of blue compact galaxies (BCG) and Galactic and extragalactic H~II regions (hereafter referred to as H2BCG; filled circles) for comparison purposes. Fig.~\ref{nevo_initial} is an analogous plot of $\epsilon(Ne)$ versus $\epsilon(O)$. In both plots the abundances for the  Milky Way PN sample were taken from HKB04, \citet{mkhs}, and \citet{hk10}, where data for all objects were obtained, reduced, and measured by the same team of astronomers. In addition, all elemental abundances were recently recomputed using an updated version of the ELSA abundance package \citep{johnson06} and the results compiled into a database by Dr. Karen Kwitter and her students at Williams College. Thus, the final abundances are considered to be homogeneous. Likewise, the M31 PN abundances were taken from \citet{kwitter12}. In Fig.~\ref{nevo_initial} we have added measurements of 70 LMC PNe from \citet{ld06}. Abundances for the H2BCG comparison sample were taken directly from the literature. The BCG data were taken from \citet{it99}. For the H~II regions we used abundances in M101 from \citet{kbg03},  from \citet{shaver83} for three objects in the MWG,  and from \citet{esteban04} for the Orion Nebula. The solid line is a least squares fit to the H2BCG sample\footnote{To check on the accuracy of the H2BCG abundances, we compiled the emission line strengths employed by the various authors and recomputed the abundances using ELSA. We found the ELSA oxygen abundances were generally 1-2\% higher than the published values, while the ELSA neon and sulfur abundances were each roughly 10\% higher than the published numbers. Because of the small differences, we proceeded by using the original abundance values found in the literature.}. Solar abundances from \citet{asplund09} are shown with dashed lines. 

The difference in behavior between the PN sample and the sample of H2BCG in Fig.~\ref{svo_initial} is striking. For the latter group, the objects clearly fall along a narrow linear track which extends over roughly two orders of magnitude in metallicity. This lockstep behavior between S and O is expected, since these two elements are alpha elements, synthesized under similar conditions in massive stars. Assuming the existence of a universal initial mass function, the lockstep behavior should be galaxy-independent. As H~II region and BCG abundances represent element levels in the ISM of their host systems, their behavior exhibited in Fig.~\ref{svo_initial} is not surprising.

At the same time, PN abundances reflect not only the level of an element present in the progenitor star at the time that it formed but also any changes to those abundances brought about by nuclear processes occurring within the star during its lifetime. In the specific cases of S and O, there is currently no strong evidence, except perhaps at very low metallicity, that PN progenitor stars alter their initial levels of these two elements (see the extended discussion on this topic in section \ref{karakas} below), and so their abundances directly reflect the amounts of S and O present in the local ISM when the star formed. Under this assumption, we would expect the lockstep behavior seen in the case of the H2BCG objects to be the same for PNe. As we see in Fig.~\ref{svo_initial}, it is not.

Interestingly, the offset in S abundances between PNe and H2BCGs is not duplicated for neon abundances shown in Fig.~\ref{nevo_initial}, i.e., we see no evidence for an offset between the two object types, although PNe continue to show more scatter. Therefore the differences in PN behavior demonstrated in Figs.~\ref{svo_initial} and \ref{nevo_initial} suggest that the offset in Fig.~\ref{svo_initial} is related to sulfur and not oxygen. Analogous plots of S versus either Ne or Ar show that S is always offset from the H2BCGs by roughly the same amount and in the same direction, further emphasizing that the problem resides with sulfur. A similar conclusion is drawn when one inspects the plots of Ne, Ar, and S versus O in \citet{mkhs}.

In what follows, we carefully examine a large number of potential explanations for the sulfur anomaly. In section \ref{correl} we test for correlations between the magnitudes of the vertical offsets of PN sulfur abundances from the H2BCG track in Fig.~\ref{svo_initial}, i.e., sulfur deficits, and numerous independent parameters related to the nebula and/or central star. In section \ref{icf} we check on the robustness of the sulfur ionization correction factor (ICF) using standard photoionization models, tests of the relevance of dielectronic recombination, as well as direct measurements of the S$^{+3}$ abundance available in the literature. Section \ref{speck} looks at the possibility that S is sequestered in either dust or molecules, while section \ref{karakas} explores the possibility that the sulfur anomaly results from stellar nucleosynthesis in AGB stars. Our summary and conclusions are provided in section~\ref{summary}.

\section{Correlation Tests Involving Nebular and Stellar Properties}\label{correl}

Table~\ref{correlations} lists the results of correlation tests performed on our PN database. We define the sulfur deficit $SD \equiv \epsilon(S_{exp}) - \epsilon(S_{icf})$, where $\epsilon(S_{exp})$ is the expected S abundance, i.e. the abundance that a PN of a given O abundance would possess if its S abundance placed it on the H2BCG track, while $\epsilon(S_{icf})$ is the S abundance of a PN inferred with the use of an ICF. Thus the more that a PN is displaced below the H2BCG track the more positive the $SD$ value. For each test the sulfur deficit was plotted against one of the 24 separate independent parameters listed in column~1 for the number of objects in column~2. The variables tested, from top to bottom in column~1, are: reddening factor, log of the ion ratio, oxygen ionization correction factor, sulfur ionization correction factor, log of the nebular diameter, central star mass, log of the central star's luminosity, neon deficit (defined in the same way as the sulfur deficit), log oxygen abundance, log of the ion ratios (4 rows), log of the difference between a PN's measured O abundance and the ambient interstellar abundance at its location along the disk, the PN's galactocentric distance along the disk, electron temperatures derived from the lines indicated (next five variables), electron densities based on the lines indicated (next two variables), and excitation class based upon the indicated emission lines and employing expressions for excitation class developed by \citet{dopita90}.

Column~3 displays the value of the Pearson correlation coefficient, while column~4 lists the correlation probability, i.e., the probability that the correlation could have resulted by choosing the same number of objects from a completely uncorrelated parent population\footnote{We employed the {\it pearsn} subroutine found in \citet{press03}}. Defining r=0.6 as the minimum value for a likely correlation (our justification of this is explained in Appendix~A), all but one of the tests indicate the absence of a correlation. The exception is the relation between the SD and $\log(S^{+3}/S^{+2})$. The increase of this ratio with the SD suggests that the deficit is due in part to the failure of sulfur ICFs to correctly account for the higher ionization stages of sulfur. Interestingly, we see no significant correlation between the SD and either the $Ar^{+4}/Ar^{+3}$ or $Ar^{+3}/Ar^{+2}$ ratios.  The ionization potentials of S$^{+2}$ and S$^{+3}$ are 34.8~eV and 47.2~eV, while those of $Ar^{+2}$ and $Ar^{+3}$  are 40.7~eV, and 59.8~eV, respectively. Since we see no significant correlation for  $Ar^{+3}/Ar^{+2}$, this suggests that there should be little sulfur in stages above S$^{+3}$. If the sulfur anomaly is the result of underestimating the population of ionization stages above S$^{+2}$, it appears that the only significant stage in this case would be S$^{+3}$. We conclude that, with the exception of $\log(S^{+3}/S^{+2})$, none of the other factors listed in Table~\ref{correlations} is significantly related to the sulfur anomaly. We should point out, however, that the question of a link between sulfur deficit and PN morphology was not explored. Such a study should perhaps be undertaken in the future.

\section{Evaluating the Effectiveness of the Sulfur Ionization Correction Factor}\label{icf}

The ionization correction factor is the ratio of the elemental abundance to the total of the observed ionic abundances. It corrects for the unobservable ions of an element. In the case of sulfur, normally only the S$^+$ and S$^{+2}$ ions are directly observable, although we know from photoionization models that higher stages, e.g., S$^{+3}$, are present in PNe in significant quantities. ICFs are usually based upon nebular models from which the abundances of the unseen ions can be estimated. It is therefore natural to speculate that the observed PN sulfur deficits are due to a formulated ICF for sulfur which incorrectly adjusts for the unseen ions.

Various formulas for the sulfur ICF have been developed over the last four decades, including those of \citet{pc69}, \citet{stasinska78}, \citet{french81}, \citet{kb94}, and \citet{kwitter01}. With the advent of ISO and Spitzer, however, the S$^{+3}$ ion has been observed directly in over 30 PNe in which oxygen abundances were also reported by \citet{pbs10}. These observations have lessened the need of an ICF in these PNe and allowed for the more direct computation of the S abundance, although the sulfur deficit problem has not been discussed extensively in the context of these observations.

In this section we check to see if photoionization models, from which ICFs are derived, reproduce the sulfur deficits observed in three sample PNe. We also test the relevance of dielectronic recombination and the possibility that incorrect rates could produce flawed ICFs. Finally, we report on the implications that direct IR measurements of S$^{+3}$ have for the sulfur deficits. 

\subsection{Photoionization Models}

\subsubsection{Standard Models}

The abundance analysis by HKB04 of their PN sample included the calculation of photoionization models of three PNe representing a large range in the sulfur deficit, i.e., IC~4593, Hu~2-1, and NGC~3242. The program CLOUDY \citep{ferland98} was used along with input blackbody stellar SEDs to model each object's measured line strengths. In the process, elemental abundances, which comprised part of the model input information, were adjusted along with other parameters such as gas density and central star temperature and luminosity in order to produce a list of predicted line strengths which closely matched the observed values. [Since then, the sulfur abundance of NGC~3242 has been revised and a new model of that PN was recently calculated by Matthew Hosek at Williams College using a central star SED of 91,000~K from \citet{rauch97}. Results of the new model are used in this discussion.] Table~\ref{models} provides a comparison of empirical and model O and S abundances along with associated sulfur deficits. We see that the model abundances for O and S along with the resulting sulfur deficit agree well with the analogous values inferred from observations for each of the three PNe used in this analysis.

Clearly the fact that models and observations agree closely suggests that either standard models should not be used to derive sulfur ICFs, or the sulfur ICF is correct and there is a real gas-phase sulfur abundance deficit in these nebulae. In the former case, some S may be sequestered in ionization stages whose populations are underestimated by the models for reasons that are unknown at this time. Next, we consider possible effects of dielectronic recombination on the ionization structure of S in model nebulae, and by extension, the ICF for sulfur.

\subsubsection{Dielectronic Recombination}

Many of the sources of atomic data used in the photoionization code CLOUDY are summarized in \citet{ferland98} with updates on the Cloudy web site www.nublado.org.  Could uncertainties in the atomic data produce the sulfur anomaly discussed here?  

The ionization balance of hydrogen and helium are quite reliable, since the ionization and recombination of these elements is dominated by valence shell photoionization and radiative recombination.  These processes are discussed, for instance, in \citet[Chapter 2]{osterbrock06}. 

The situation is complex and more uncertain for the heavy elements.  Ionization for a PN is predominantly by valence shell photoionization, although inner shell photoionization and Auger decay can be significant \citep[Chapter 11]{osterbrock06}.  The heavy elements primarily recombine through dielectronic recombination, a process where a free electron is captured into autoionizing levels.  These are levels with two excited electrons and with energies above the ionization limit of the recombined atom \citep[Appendix A4]{osterbrock06}.  These rates of dielectronic recombination (DR) are sensitive to the energies of the autoionizing levels because they determine whether free electrons can reach those levels.  In practice, accurate DR rates require experimental autoionization level energies.  

For high gas temperatures, we use the recombination data set collected by Nigel Badnell for both radiative and dielectric recombination\footnote{These rates are available at the following websites: http://amdpp.phys.strath.ac.uk/tamoc/DR/ (dielectronic recombination) and http://amdpp.phys.strath.ac.uk/tamoc/RR/ (radiative recombination).}. DR rates are available for elements of the second row (O and Ne are important for optical spectra of PN) but only for relatively high ions of the third row. In particular, experimental energies are not available for the low ions of sulfur studied in this paper, and their sulfur DR rates are not given by Badnell.  Instead, for these ions we use estimates of the rates based on the simple averaging approach described by \citet{ali91}. That is, the Badnell rates are joined on to the mean of charge-specific CNO rates at lower temperatures.   

It is impossible to say how uncertain these estimates might be.  Comparing with other ions this method generally works to within a factor of two.  Could such uncertainties affect the ionization balance and line emission enough to cause models to misrepresent what happens in real nebulae?  We vary the estimated DR rate by these amounts to find out.

In the present case, we investigated the effects of varying the DR rate on sulfur emission line strengths as a way of identifying a mechanism which could profoundly affect the ICF of sulfur in a way that would explain the observed sulfur deficits. For example, overestimated rates would affect the sulfur ionization balance by shifting sulfur into lower stages, causing an ICF based on these models to underestimate the sulfur elemental abundance. An opposite shift would result if rates used are below their actual values. 

We therefore computed three grids of photoionization models using CLOUDY 08.01, where input parameter ranges for each grid were designed so that the output predicted line strengths spanned the space of observable line ratios for the PN sample. Three model input parameters were varied: the SED of the central star, the overall metallicity, and the nebular gas density. For stellar temperatures between 30,000 and 50,000~K we used the Atlas SEDs by  \citet{kurucz91}, while for temperatures above 50,000 K we employed the SEDs of \citet{rauch97}. For each grid, input stellar temperature ranged from 30,000-158,000 K, total gas density from 10$^2$-10$^5$, and metallicity from 0.25-2.5 times solar. The DR rates in the first grid were the same as those prescribed by \citet{badnell91}. For the other two grids, the rates were halved or doubled. All models were radiation bounded.

\begin{figure}[h]
   \includegraphics[width=6in,angle=270]{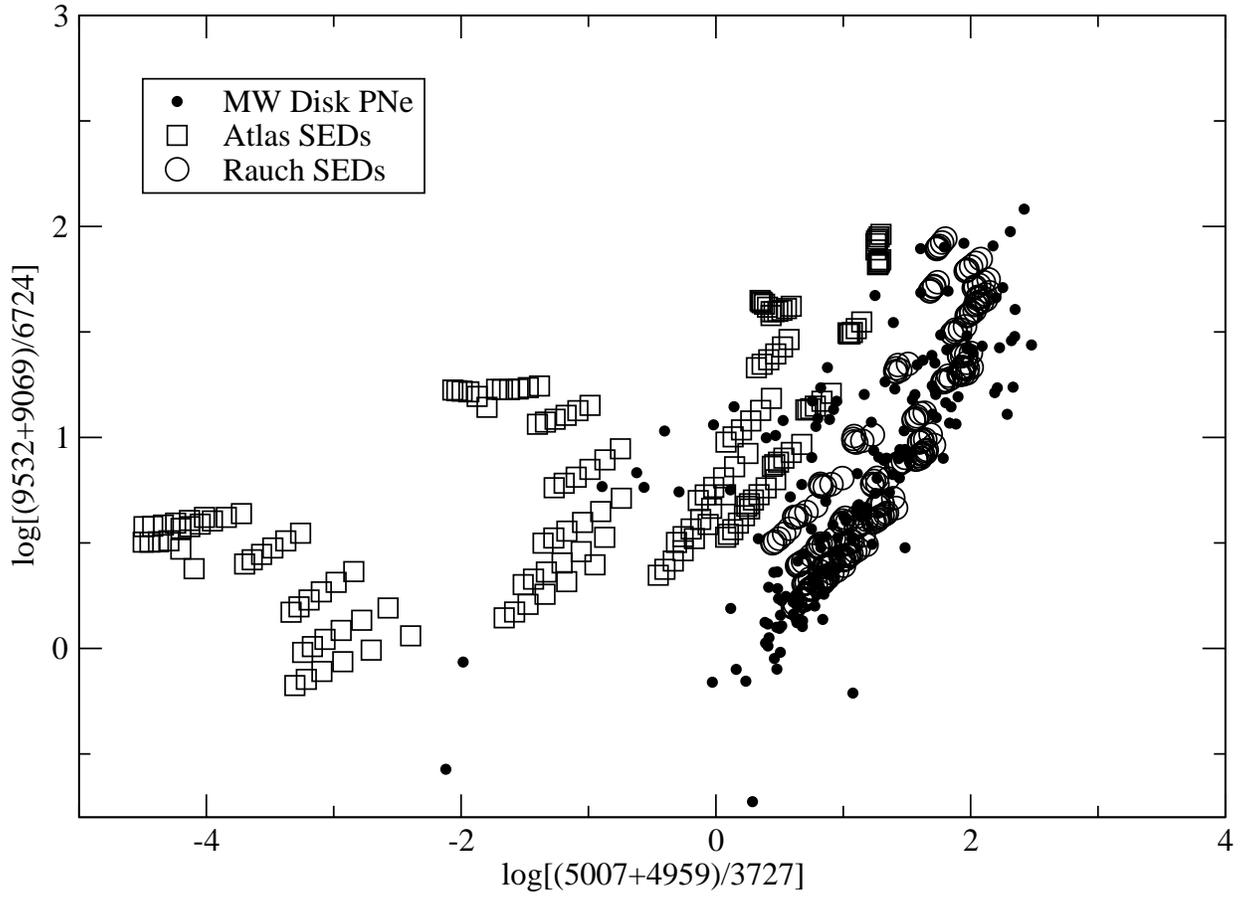}
   \caption{Logs of the [S~III]/[S~II] and [O~III]/[O~II] ratios plotted against each other for the same observed MW disk PN sample displayed in Fig.~\ref{svo_initial}  
   (filled circles) and photoionization models using Atlas (open squares) or Rauch AGB (open circles) input SEDs.}
\label{dr}
\end{figure}

Results for the first grid are shown in Fig.~\ref{dr}, where for each model we computed the predicted line strength ratio of ([S~III] $\lambda$9523+$\lambda$9069)/[S~II] $\lambda$6724 and ([O~III] $\lambda$5007+$\lambda$4959)/[O~II]$\lambda$3727. Predicted ratios for models employing the Atlas or Rauch SEDs are shown with open squares and circles, respectively, while the line ratios for the same MW disk PN sample shown in Fig.~\ref{svo_initial} are shown here with filled circles. First, we note that these models mostly span the space occupied by the observations. The exceptions appear to be the objects with low [S~III]/[S~II] ratios along with a few with high [O~III]/[O~II] ratios. In the first case, the observed reduced sulfur ratios are likely the result of partial telluric absorption of the near IR lines at $\lambda\lambda$9069,9532. The second group is more difficult to characterize. However, in comparing a number of their observed properties with those of the majority of the sample, all members of the high [O~III]/[O~II] group are likely to be matter bounded. For example, compared with the rest of the sample, these objects all possess relatively low observed values of [O~I] $\lambda$6300, [O~II] $\lambda$3727, and [N~II] $\lambda$6584, i.e., lines typically reduced in strength relative to those produced by higher ionization stages when the nebula is matter bounded.

\begin{figure}[h]
   \includegraphics[width=6in,angle=270]{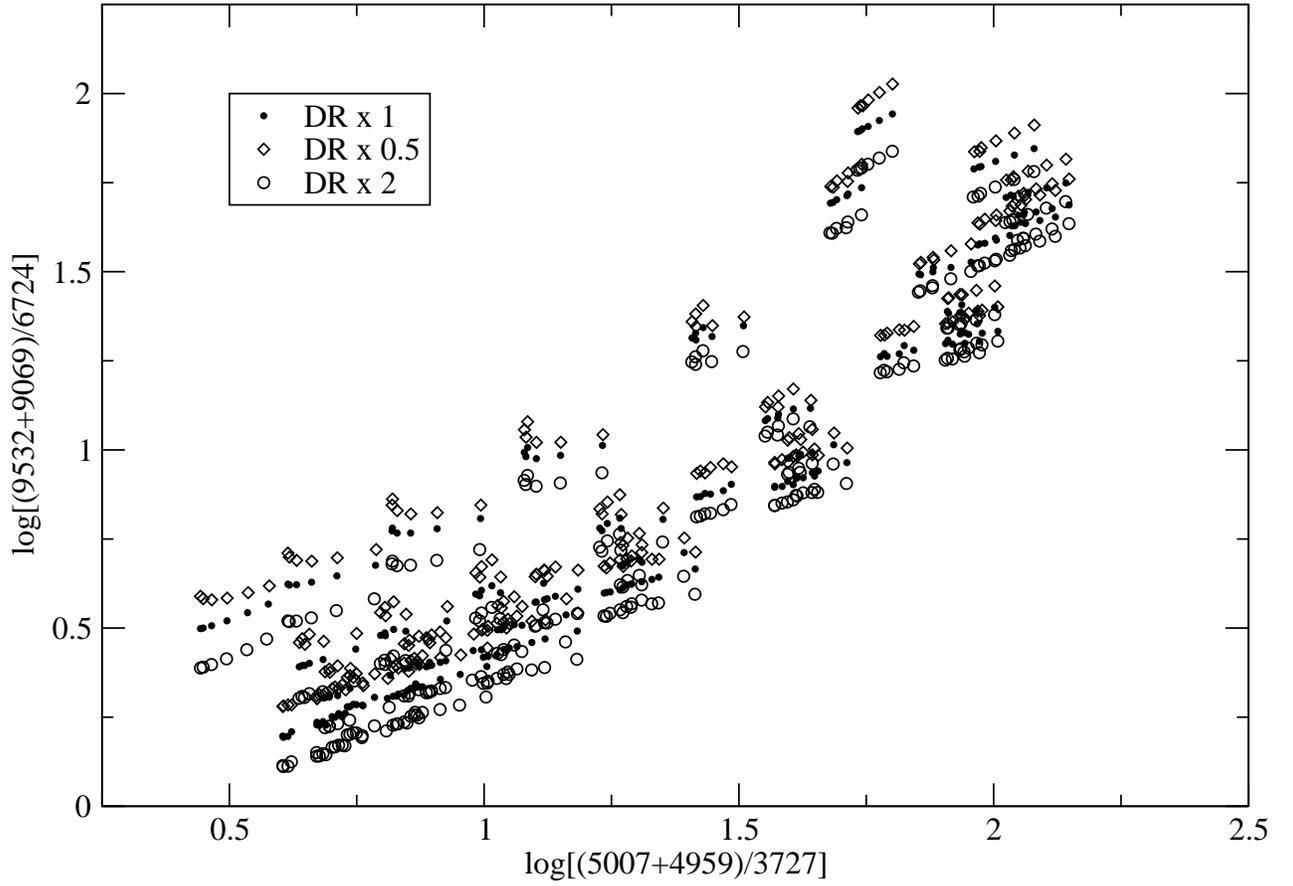}
   \caption{Logs of the [S~III]/[S~II] and [O~III]/[O~II] ratios plotted against each other for photoionization models only, using the standard DR rates (closed circles), one-half the standard rates (open diamonds), and twice the standard rates (open circles).}
\label{dr_rauch}
\end{figure}

Figure~\ref{dr_rauch} demonstrates the effects of multiplying the Badnell DR rates by factors of 0.5 and 2. Here we plot results for the same line ratios as in Fig.~\ref{dr} for all three grids. Only the models employing the Rauch AGB SEDs are shown. The figure clearly shows the effects of changing the rates. Typically the model points shift up by about 0.1~dex when the rate is halved and down by a similar amount when the rate is doubled, the directions of the shifts being in the sense that is expected. The thermostat effect, where changes in the abundance of an ion do not linearly affect its emission lines due to energy balance, accounts for the relatively modest change in line emission, despite large changes in the assumed DR rates. Shifts are similar for those models based upon the Atlas SEDs. These changes in DR rates, representing reasonable uncertainties in their actual values, correspond to shifts in the sulfur ICF of about 13\% or 0.05~dex above or below the value obtained before DR scaling. Clearly, these small changes in the ICF cannot explain the sulfur anomaly.

\subsection{Direct Infrared Observations of S$^{+3}$}

All of the sulfur abundances shown in Fig.~\ref{svo_initial} result from applying an ICF to the observed abundances of S$^+$ and S$^{+2}$. Does replacing the ICF by directly measuring S$^{+3}$ reduce or eliminate sulfur deficits?

Sulfur abundances of 32 Galactic PNe determined through direct observations of S$^{+3}$ have been summarized by \citet{pbs10}. Their study made use of the ISO and Spitzer telescopes to measure the strength of the [S~IV] 10.5${\mu}$m line. Similar measurements using Spitzer were made by \citet{shaw10} of six PNe in the SMC and by  \citet{henry08} of DdDm-1, located in the Galactic halo. 

 In Table~\ref{ir} we provide S abundance results which have been derived using both the ICF method ($S_{icf}$) and the IR method (S$_{ir}$). For each object identified in the first column we list in columns 2-5 the values for $\epsilon(O)$; $\epsilon(S_{exp})$, i.e., the $\epsilon(S)$ value on the  H2BCG track for the given oxygen abundance; $\epsilon(S_{icf})$; and $SD_{icf}$. For the last quantity we have added the subscript to indicate that it is based on the standard ICF abundance method. Columns 6 and 7 give $\epsilon(S_{ir})$ resulting from the IR measurements and the associated sulfur deficit SD$_{ir}$ based upon values in column~3. Column 8 provides a measure of the difference in S abundances derived from the two methods. Finally, values in column 9 represent the amount of an unidentified ion of S such as S$^o$ or S$^{+4}$, relative to S$_{exp}$, needed to explain the difference between $S_{ir}$ and $S_{exp}$, i.e., $S^i/S_{exp}=1-\log^{-1}(-SD_{ir})$.
 
\begin{figure}[h]
   \includegraphics[width=6in,angle=270]{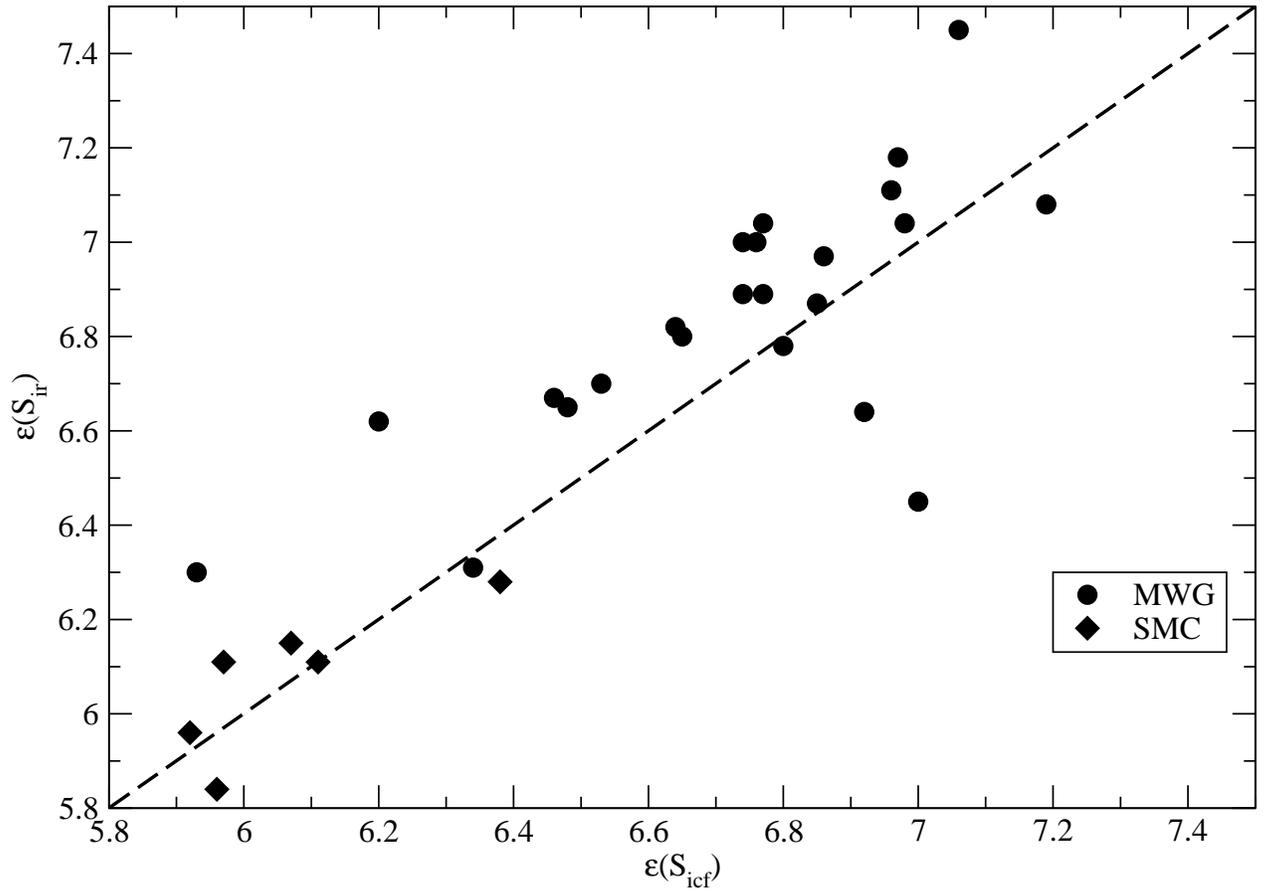}
   \caption{Sulfur abundances inferred from IR data versus S abundances derived using an ICF. Filled circles and diamonds show positions for PNe in the MWG disk and the SMC, respectively. The dashed line shows the one-to-one correspondence.}
\label{sirvsicf}
\end{figure}

Fig.~\ref{sirvsicf} shows a visual comparison of the  $\epsilon(S_{ir})$ and $\epsilon(S_{icf})$ values in Table~\ref{ir}, where Milky Way disk objects appear as filled circles, SMC objects as filled diamonds. The dashed line indicates the one-to-one correspondence. With a few exceptions there is a clear tendency (21 of 29 objects) for the IR abundances to exceed the corresponding ICF abundances. Of those 21 PNe the median positive vertical distance from the line is 0.17~dex or a factor of 1.5 times the $S_{icf}$ value. This suggests that when S$^{+3}$ observations replace the ICF method, half again as much S, on average, is inferred. This result is qualitatively consistent with the positive correlation we found in \S~2 between the SD and $\log(S^{+3}/S^{+2})$.  For those eight objects falling below the line, apparently the ICF overcorrects for unobserved sulfur ions, as would be the case if these PNe were matter-bounded.

\begin{figure}[h]
   \includegraphics[width=6in,angle=270]{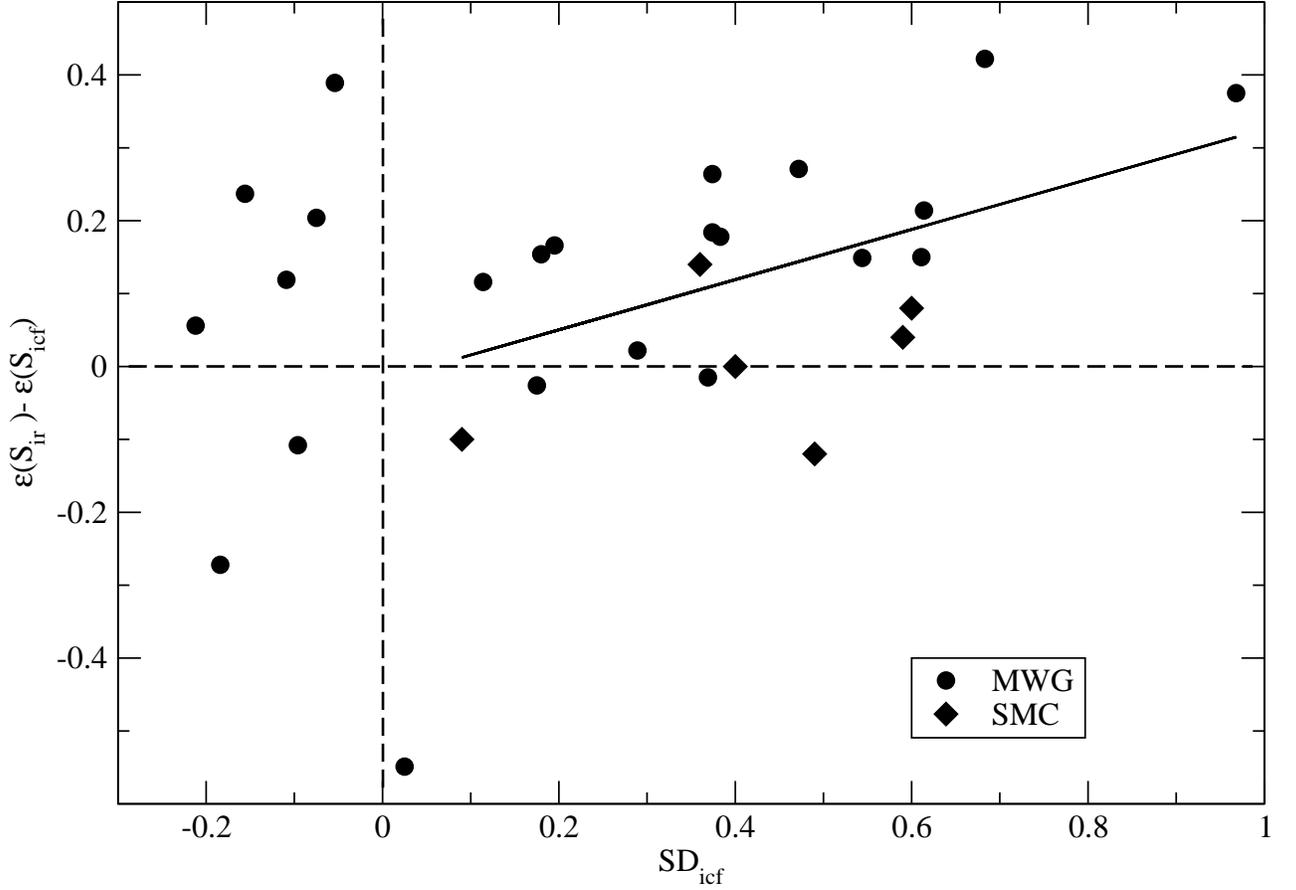}
   \caption{The difference in logarithmic S abundances inferred from IR and ICF methods versus SD$_{icf}$. Filled circles and diamonds represent PNe in the Milky Way disk and SMC, respectively. Objects to the right of the vertical dashed line have S abundances which fall below the values on the H2BCG line for their O abundance, while objects to the left of the dashed line have S abundances above that trend line. The horizontal dashed line separates the plot into regions of positive and negative values for $S_{IR} - S_{ICF}$. The solid line shows a least squares fit to the subset points with positive values for $SD_{ICF}$.}
\label{ifactor}
\end{figure}

Next, if the sulfur deficit is caused by the failure of the ICF method to properly correct for S$^{+3}$, then there should be a direct correlation between the sulfur deficit SD$_{icf}$ and the quantity $\epsilon(S_{ir})-\epsilon(S_{icf})$. Values listed in columns 5 and 8 of Table~\ref{ir} are plotted against each other in Fig.~\ref{ifactor}. For this analysis we focus on those objects with positive values for SD$_{icf}$, i.e., those to the right of the vertical dashed line, ignoring the one outlier near the vertical line at the bottom of the plot. Here we see by inspection some evidence for a positive correlation. A regression analysis of the 21 points provides the least squares fit shown by the solid line. The slope of the line is +0.34$\pm$.13 with an intercept of -0.01$\pm$.06~dex. However, the correlation coefficient $r=+0.52$, with a probability of less than 1\% that the same set of points could have come from an uncorrelated sample. As explained in Appendix~A, most statisticians require a value of $r\ge0.6$ for a true correlation, and so we conclude that the apparent trend in Fig.~\ref{ifactor} is not robust enough to confirm a relation between values in columns 5 and 8.

\begin{figure}[h]
   \includegraphics[width=6in,angle=270]{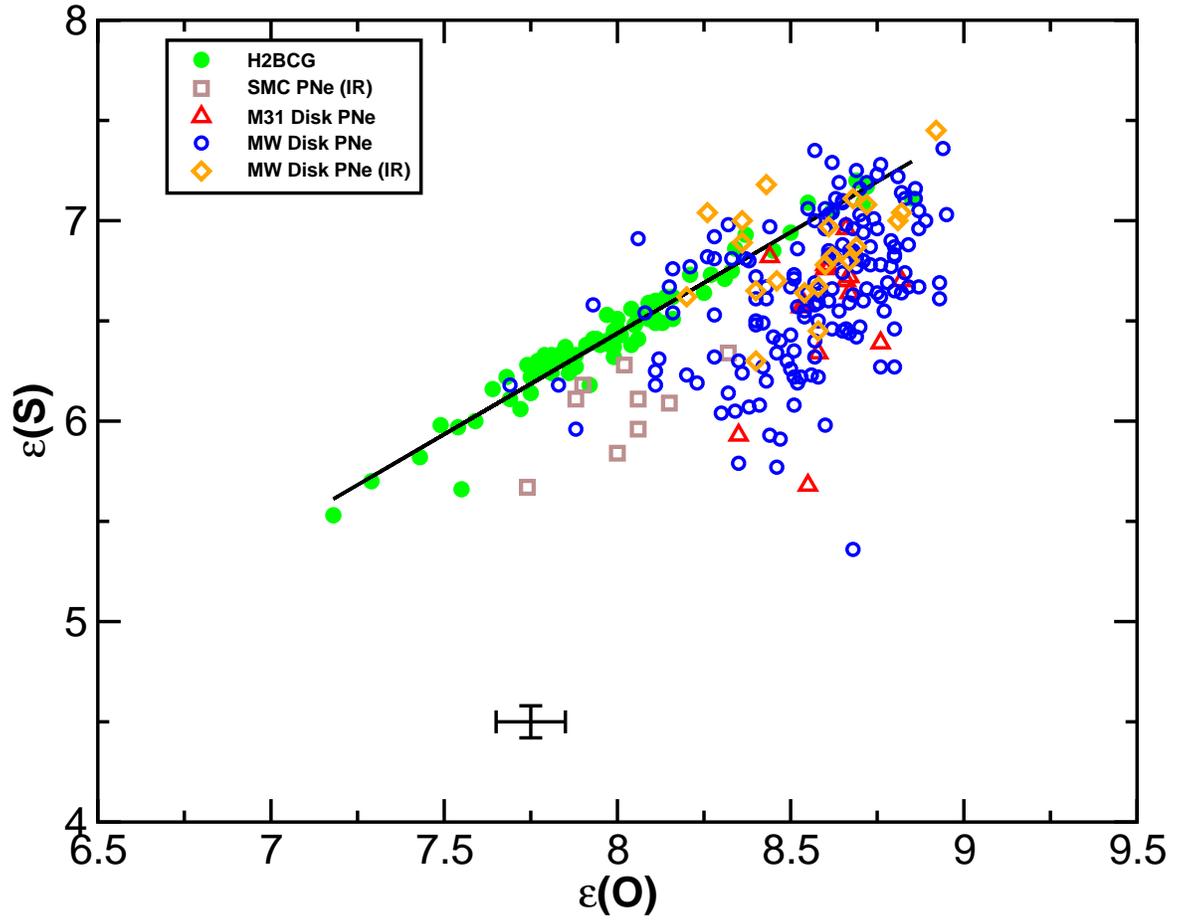}
   \caption{Same as Fig.~\ref{svo_initial} but now showing the objects from the samples of  \citet[open diamonds]{pbs10} and \citet[open squares]{shaw10} in which IR observations were used to infer S abundances.}
   \label{svo_ir}
\end{figure}

Approaching this line of investigation regarding S$^{+3}$ from another angle, however, we have added the S and O abundance data from \citet[open diamonds]{pbs10} and \citet[open squares]{shaw10} to the original plot in Fig.~\ref{svo_initial} to produce Fig.~\ref{svo_ir}. Compared with the original PN sample of open circles, we now see a reduced tendency for these objects to fall systematically below the H2BCG track. Also, the vertical scatter is much closer to what we see in the case of neon in Fig.~\ref{nevo_initial} and perhaps can be attributed simply to natural abundance scatter. Thus, the direct use of the S$^{+3}$ abundances to determine total S abundances seems to solve at least part of the problem related to the sulfur anomaly, as originally anticipated by HKB04. This finding is also consistent with the model results of \citet{goncalves11}, who use the 3-D photoionization code Mocassin to show that as the temperature of the PN central star increases, the sulfur ion population shifts upward and the ratio of the sulfur ICF of \citet{kb94} to the 'true' value inferred from the ionization populations in the model output declines.

Shifting attention now to the points in Fig.~\ref{ifactor} with negative $SD_{icf}$ values, we see that there is no apparent relation between the parameters plotted. If we speculate that the sulfur anomaly is the result of an unknown nebular property that populates unseen ionization stages of S beyond what we find in the models discussed above, then we can conclude that this same property is not relevant for these seven objects. For them, other factors are influencing the values of $S_{ir} - S_{icf}$, but they appear to be unrelated to the sulfur anomaly.

We turn now to the interpretation of the values posted in the last column in Table~\ref{ir}, i.e., $S^i/S_{exp}$ [$=1-\log^{-1}(-SD_{ir})$]. For a given oxygen abundance, if we take S$_{exp}$ to be the sulfur abundance of a PN falling on the H2BCG track, then the value in column 9 is the fraction of that total abundance that represents the difference between it and the sum of S$^+$, S$^{+2}$, and S$^{+3}$ measurements, i.e., S$_{ir}$. (The six objects with negative $SD_{ir}$ values are irrelevant here.) The average value of $S^i/S_{exp}$ is 0.43$\pm$.24. This is rougly the level of the estimated natural scatter inferred from neon abundances of about 0.2~dex or 60\% (see Fig.~\ref{nevo_initial}).  Generally, then, replacing the ICF with direct S$^{+3}$ measurements appears to reduce sulfur deficits in most PNe to the level of expected scatter in the S abundance. However, we note that these objects still fall systematically below the H2BCG track, and thus the deficits of PNe are not entirely explained by employing the direct measurement of S$^{+3}$. This conclusion is consistent with results reported by \citet{bs06} and \citet{pbs06}. 

Values in column~9 surpassing this scatter level would suggest that additional ions of S such as S$^o$ and/or S$^{+4}$, are present in significant amounts in these objects. We note that an unexpectedly high abundance of ions such as S$^{+4}$ in PNe is predicted by nebular models by \citet{jacob11} in which the density profile of the nebula is non-uniform. Our standard CLOUDY models suggest that S$^{+4}$ has a permitted line S~V $\lambda$1198 corresponding to the 3s3p $^3$P to 3s$^2$ $^1$S$_0$ transition. If we assume an ion fraction of 10\% for S$^{+4}$, then, extrapolating from these same model predictions, the S~V line will be well below detection levels. Note, however, that considerations of ion ratios and ionization potentials in \S\ 2 appear to conflict with the idea of significant abundances of ions above S$^{+3}$.

\section{Dust and Molecules}\label{speck}

In studies of the ISM, deficiencies in the abundances of elements in atomic
(or ionized) gas is usually assumed to be due to depletion into dust grains
\citep[e.g.][]{whittet1992,krugel2008}.
Consequently, the observed sulfur deficit in PNe has been attributed
to sequestration into dust grains \citep{jero}.
%
%
The chemical composition of dust in PNe should reflect that of their
parent Asymptotic Giant Branch (AGB) stars.
The atmospheres of AGB stars are expected
to change as these stars evolve, due to convective dredge up of carbon
produced in the He-burning shell. The amount of carbon relative to oxygen (the
C/O ratio) is critical in determining which types of dust and molecules
are present around
an AGB star. The formation of extremely stable CO molecules will consume
whichever of the two elements is less abundant, leaving only the more abundant
element available for dust formation. Stars start their lives with the cosmic
C/O ratio of $\approx$0.4 and are therefore oxygen-rich. In about a third of
AGB stars, enough carbon will be dredged up to make C/O $>$ 1 and therefore
carbon will dominate the chemistry around these stars.
%

Sulfur-bearing solids (e.g. MgS, CaS and FeS) are predicted to
form in carbon-rich environments \citep[e.g.][]{nuth1985,lodders1995}
and have been tentatively observed in
carbon-rich AGB stars, PPNe and  PNe, e.g. the ``30\,$\mu$m'' feature
oftentimes associated with MgS
\citep[e.g.][]{Goebel1985,nuth1985,Omont1995,Begemann1994,Hony2002}.

\begin{figure}[h]
   \includegraphics[width=5.5in,angle=270]{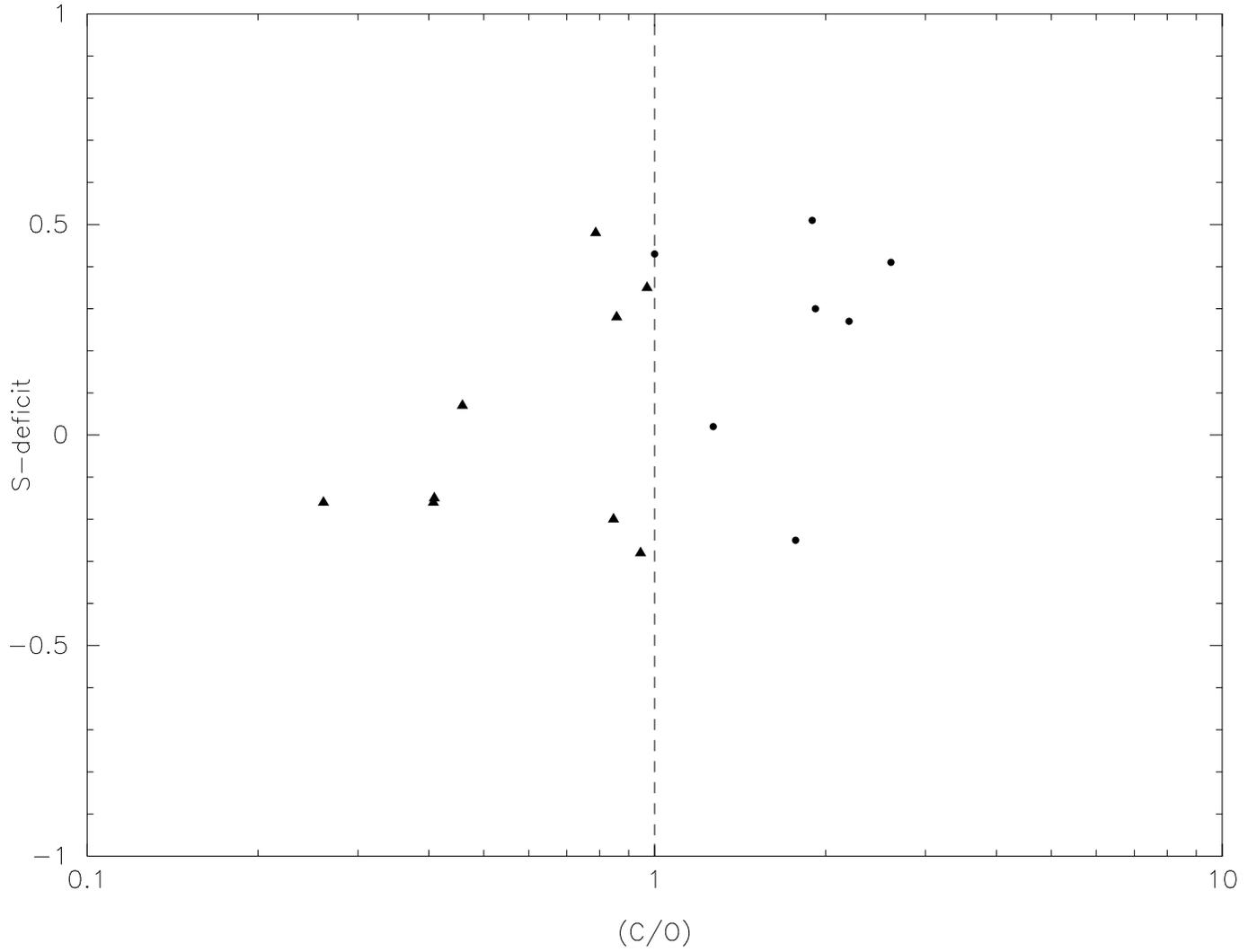}
   \caption{Sulfur deficit versus the observed C/O abundance ratio for those sample objects for which carbon has been measured. The vertical dashed line divides the space between C/O$<$1 (oxygen-rich; triangles) and C/O$>$1 (carbon-rich; circles) environments.}
\label{covsd}
\end{figure}

\begin{figure}[h]
\includegraphics[angle=270,scale=0.6]{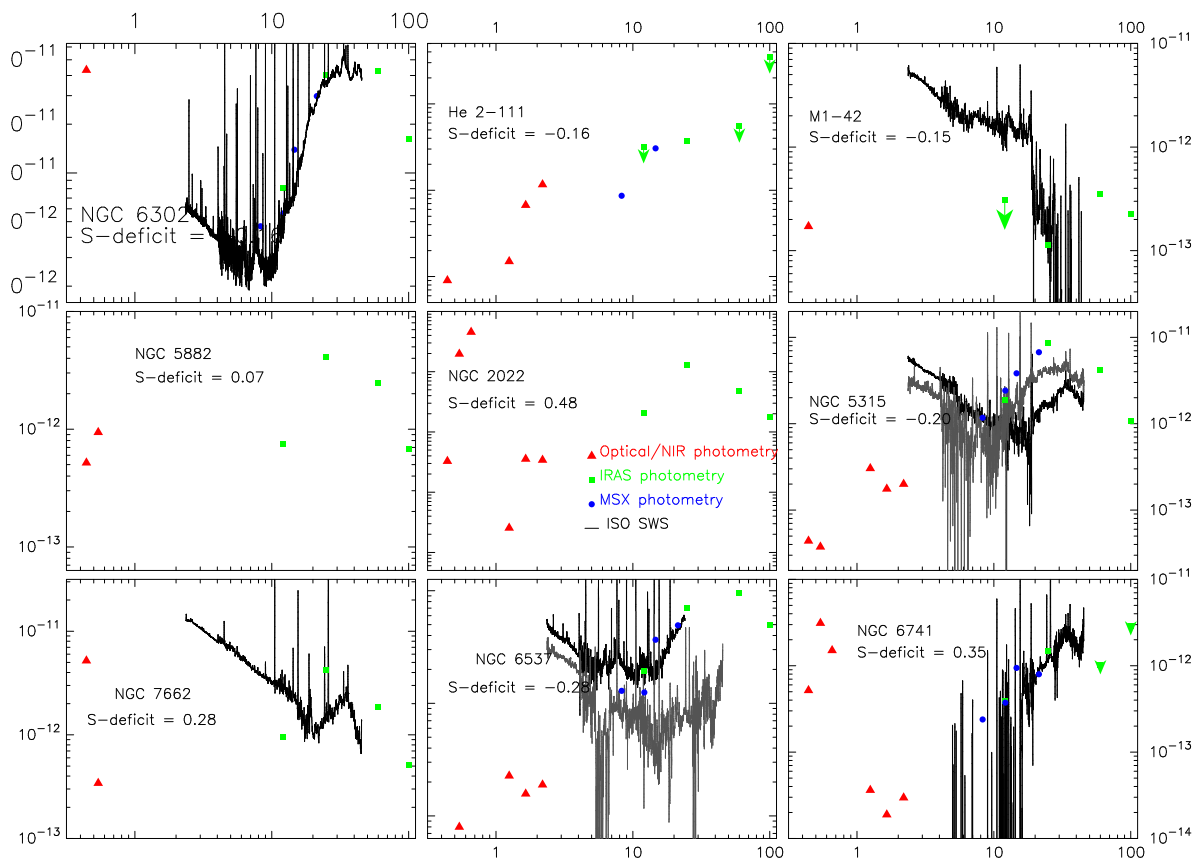}
\caption{\label{dustemissionO}
Spectral energy distributions for oxygen-rich PNe. $x$ axis is wavelength in $\mu$m;
$y$ axis is $\lambda$\,$F_\lambda$ in W\,m$^{-2}$;
filled triangles are the optical and near-IR photometry (collected from Simbad);
filled squares are IRAS photometry;
filled circles are MSX photometry; and
solid line is the ISO SWS spectrum.
Arrows indicate an upper limit on the photometry points.
The S-deficit is indicated in the legend for each SED.}
\end{figure}

\begin{figure}[h]
\includegraphics[angle=270,scale=0.6]{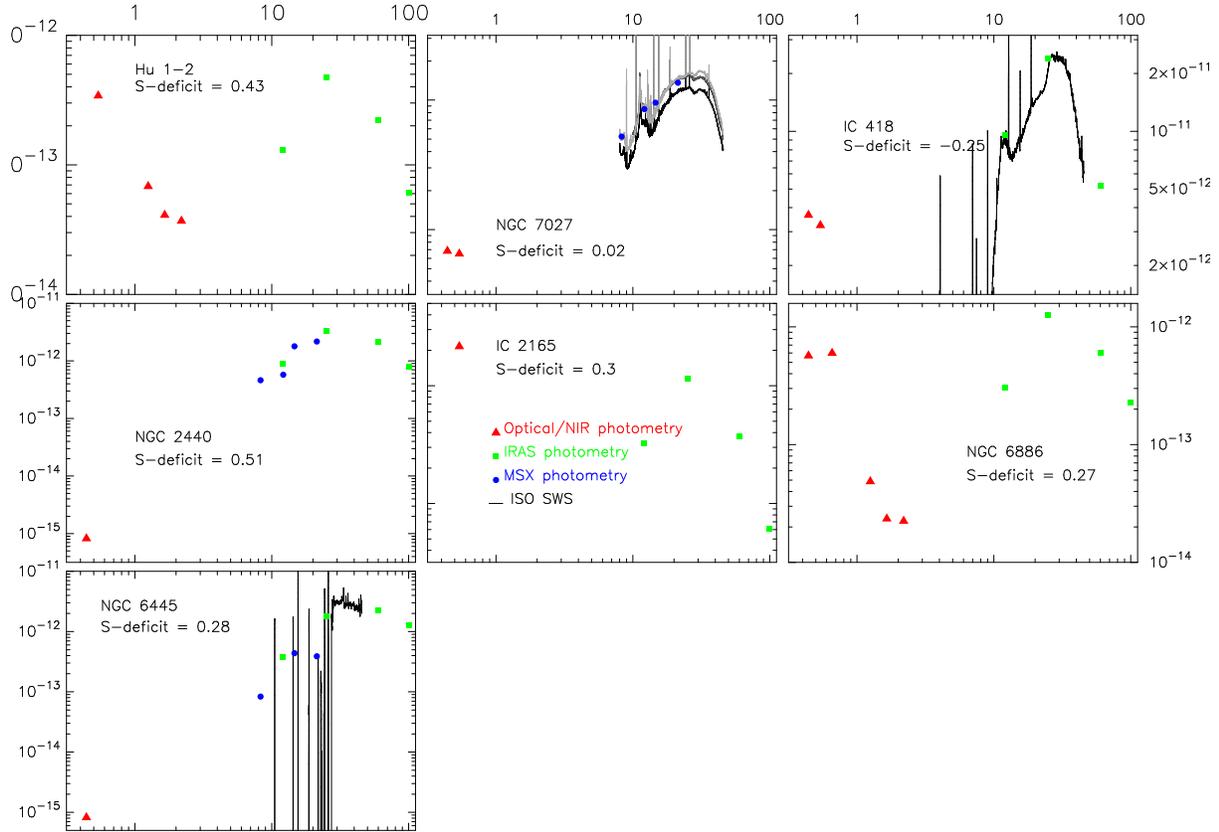}
\caption{\label{dustemissionC}
Spectral energy distributions for carbon-rich PNe.} $x$ axis is wavelength in $\mu$m;
$y$ axis is $\lambda$\,$F_\lambda$ in W\,m$^{-2}$; filled triangles are the optical and near-IR photometry
(collected from Simbad);
filled squares are IRAS photometry;
filled circles are MSX photometry; and
solid line is the ISO SWS spectrum.
Arrows indicate an upper limit on the photometry points.
The S-deficit is indicated in the legend for each SED.
\end{figure}

In oxygen-rich environments sulfur-rich grains are
precluded by the abundant free oxygen, which will tend to oxidize any
sulfides that form, releasing the sulfur back to the gas phase. Therefore,
we should expect sulfur depletion in the gas phase to be highest in
carbon-rich nebulae. Fig.~\ref{covsd} shows that the distribution of sulfur
deficits does not correlate with the the carbon abundance
(the determination coefficient r$^2$ is 0.202 or r=0.45; see Appendix~A for a brief discussion of correlation coefficients.)
The figure shows that oxygen-rich PNe
possess the same spread in sulfur deficits as their carbon-rich counterparts.
Furthermore, there is no correlation between the sulfur deficit and the
occurrence of the ``30\,$\mu$m'' feature; some carbon-rich PNe which exhibit
 the ``30\,$\mu$m'' dust spectral feature
also show low sulfur deficits or even enhanced sulfur abundances.
Figures~\ref{dustemissionO} and \ref{dustemissionC} show the spectral energy
distributions for a selection of oxygen-rich and carbon-rich PNe, respectively.
Although ISO SWS spectra are not available for all objects, it is clear that,
for instance, NGC~7027 exhibits a ``30\,$\mu$m'' feature but is barely
sulfur-deficient (SD = 0.02); while IC~418, which is enriched in sulfur
(SD = $-$0.25),  also has a strong ``30\,$\mu$m'' feature.

\begin{figure}[h]
\includegraphics[angle=270,scale=0.6]{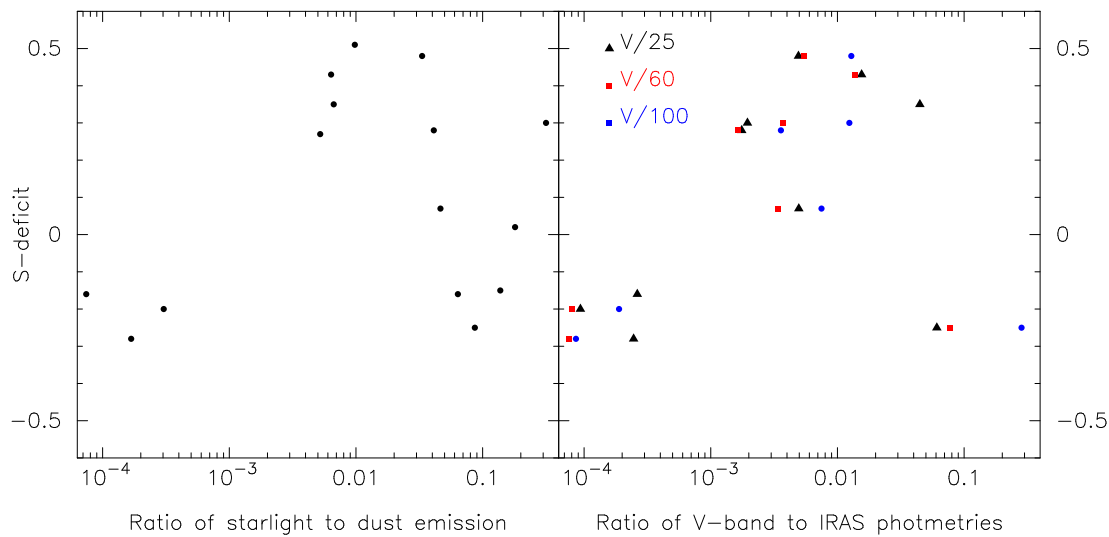}
\caption{\label{ratio}
The ratio of starlight to dust emission as a function of sulfur deficit.
{\em Left panel} the ratio of starlight to dust emission is determined by
calculating the area under each hump in the SED for each nebula.
{\em Right panel} the ratio of starlight to dust emission is estimated using the V-band photometry as a proxy for starlight, and the IRAS 25, 60 and 100\,$\mu$m photometries as proxies for dust emission.
In both cases $x$-axis is the ratio of starlight to dust emission;
the dustiness of the nebulae increases to the left.
$y$-axis is the S-deficit.}
\end{figure}

If sulfur-bearing dust does not exhibit spectral features
(only contributes to continuum) and
can exist in an O-rich environment, we would still expect to see a correlation
between the dust abundance and the sulfur deficit. This is not the case.
PNe typically show double-humped SEDs, with a short-wavelength hump due to starlight and a long-wavelength hump due to dust emission.
Examination of the SEDs presented in
Figures~\ref{dustemissionO} and \ref{dustemissionC}
show that there is no correlation between the
S-deficit and whether
starlight or dust emission dominates the SED.
In order to quantify the ratio of starlight to dust emission we have
determined the flux contained in the starlight and dust emission humps by
interpolating between the photometry points and calculating the flux beneath
each hump. The ratio of starlight to dust emission as a function of
sulfur-deficit is plotted in Figure~\ref{ratio}.

In some cases, the photometry measurements available are sparse
(e.g.\ IC~2165), which makes the determination of the starlight to dust
emission ratio uncertain. Consequently Figure~\ref{ratio} also contains the
ratio of the V-band photometry to that in the infrared for the 25, 60 and
100\,$\mu$m IRAS photometry - as these far-IR photometry points should be
good proxies for dust emission strength.
We cannot find any correlation between dust emission and the sulfur-deficit
and thus we rule out dust
formation as the source of the sulfur abundance anomaly.

Another potential sink for
atomic/ionized sulfur gas is molecule formation. Like dust grains, molecules
form during the
AGB phase. Sulfur-bearing molecules observed around AGB stars include
SO, SO$_2$, H$_2$S, CS and SiS \citep{omont1993,bujarrabal1994},
and can exist in both C-rich
and O-rich environments.
As the star evolves towards the PN phase, the molecules will be
destroyed by the increasingly high energy photons emanating from the star.
However, molecules can survive in highly evolved PNe
\citep[e.g. the Helix Nebula;][and references therein]{matsuura2009}
as long as they are shielded from the high energy photons by dust (possibly
in clumps). Sulfur-rich molecules will not return their sulfur bounty to the
gas if the dust shields them. In this case we would still expect to see a
correlation between the dust abundance and the sulfur depletion. As shown
above (Figures~\ref{dustemissionO} and \ref{dustemissionC}),
this is not the case. However, dust emission in clumpy
nebulae (like the Helix Nebula) is weak even though the presence of dust is
inferred from the survival of molecular species.

\begin{figure}[h]
\includegraphics[angle=270,scale=0.6]{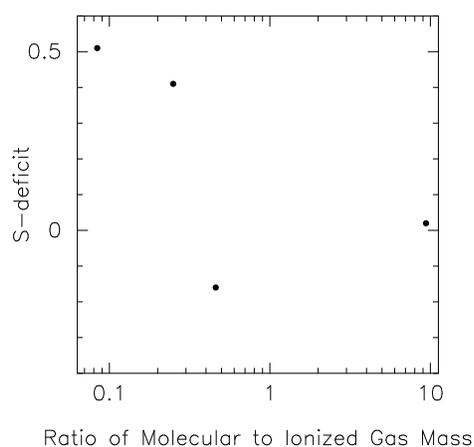}
\caption{\label{molecules}
The ratio of the molecular gas mass to ionized gas mass as a
function of sulfur deficit.}
\end{figure}

In order to test the hypothesis that sulfur can be sequestered in molecules in
clumpy nebulae we sought correlations between the ratio of ionized/molecular
mass in PNe with the sulfur deficit (Figure~\ref{molecules}). The ratios of
molecular to ionized gas are taken from \citet{huggins1,huggins2}.
Unfortunately,
the number of PNe for which we have data on both molecular mass and sulfur
anomolies is too small to be statistically significant. However, the limited
sample in Figure~\ref{molecules} suggests that the highest S-deficit is
associated with the lowest molecule content.

We have shown that sulfur abundance anomalies cannot be attributed to
sequestration of sulfur into dust grains. Formation and protection of
sulfur-bearing molecules may provide a mechanism for sulfur sequestration.
Further study of correlations between sulfur abundances and clumpiness of the
host nebulae, and with evolutionary status of the nebulae are required.

\section{Nucleosynthesis in Asymptotic Giant Branch stars}\label{karakas}

The illuminated PN is comprised of material
from the deep convective envelope, hence nebular abundances should
reveal information about the efficiency of mixing events and chemical
processing that took place during previous evolutionary phases,
in addition to the initial composition of the parent star
\citep{dopita97,karakas03a,karakas09}.  In light of this information,
is it possible to explain the sulfur anomaly in PNe by
examining the observed or predicted nucleosynthesis outcomes from
AGB stars?
We start with a brief review of AGB evolution and nucleosynthesis.

During the TP-AGB phase the He-burning shell becomes
thermally unstable every $10^{5}$ years or so,
depending on the core mass. We refer to \citet{busso99} and
\citet{herwig05} for reviews of AGB evolution and
nucleosynthesis. The energy from the thermal pulse
(TP) drives a convective pocket in the He-rich intershell,
that mixes the products of He-nucleosynthesis within this region.
The energy provided by the TP expands the whole star, pushing
the H-shell out to cooler regions where it is almost
extinguished, and subsequently allowing the convective
envelope to move inwards (in mass) to regions previously
mixed by the flash-driven convective pocket.
This inward movement of the convective envelope is known as
the third dredge-up (TDU), and is responsible for enriching
the surface in $^{12}$C and other products of He-burning,
as well as heavy elements produced by the $slow$ neutron
capture process (the $s$ process). Following the TDU, the star
contracts and the H-shell is re-ignited, providing most of
the surface luminosity for the next interpulse period.
In intermediate-mass AGB stars with initial masses
$\gtrsim 4 M_{\odot}$ (with core masses $\gtrsim 0.8M_{\odot}$), the
base of the convective envelope can penetrate into the top of the
H-burning shell, causing proton-capture nucleosynthesis to
occur there (hot bottom burning, HBB). The surface composition
of the star will be changed owing to the fact that
the entire envelope is exposed to the hot
burning region a few thousand times per interpulse period. Given
the fast evolutionary timescales of post-AGB stars with
$M_{core} \gtrsim 0.8 M_{\odot}$ \citep{bloecker91} it is not clear how
many end up illuminated as PNe. For this reason, we will
concentrate on nucleosynthesis in lower mass AGB stars.

In low-mass AGB stars the
evolution of the composition of the surface layers
is primarily affected by the occurrence and efficiency
of the TDU.  The efficiency of the TDU and the composition of the
He-intershell have been shown to vary as a function of the
initial mass, metallicity, as well as time
during the AGB \citep{boothroyd88d,karakas02,straniero03,karakas10a}.
Observationally these quantities are not well known but a few
general trends can be inferred.  At the metallicity of the Galactic disk,
the TDU likely occurs in stars with initial masses
$\gtrsim 1.5 M_{\odot}$; this mass is reduced
to $\sim 1 M_{\odot}$ at the metallicities of the Magellanic Clouds
[with metallicities $\sim 2-5$ times less than solar \citep{wallerstein98}].
Of particular interest
for this study is the nucleosynthesis that occurs in the He-intershell.
This is harder to determine observationally. The subclass of PG1159
post-AGB stars are H-deficient and show what has been considered
to be He-intershell material at their surface \citep{werner09}.
In these stars, C abundances vary from 15--60\% (by mass),
and O from 2--20\%. There are also trace quantities of other He-shell
burning products including Ne ($\sim$2\%, probably in the form
of $^{22}$Ne) and F. Phosphorous is solar, sulfur has abundances
varying from 0.01 times solar to solar, and some PG 1159 stars
are also highly deficient in Fe \citep{werner06}.

The C and O PG 1159 abundances are in direct contrast
to intershell abundances predicted by stellar evolution
models of AGB stars \citep[e.g.,][]{karakas10a}.
Standard models predict that He
constitutes about 75\% (by mass), C roughly $\sim 25$\%,
and there is about $1-2$\% each of  $^{22}$Ne and $^{16}$O.
The predicted abundances of F and Ne agree with the abundances
measured in PG 1159 stars \citep{werner94,werner05}.
Ne from the intershell is predicted to increase the surface
Ne/O ratio, but only when there is very efficient TDU and this
has been predicted to occur in only a narrow mass range
\citep[$\sim 2.5-3.5 M_{\odot}$][]{karakas03a,karakas09}.
The remarkably constant Ne/O ratio observed in PNe with a wide
variety of chemical compositions indicates that in most
cases this ratio is not altered during the AGB. 
This is consistent with the conclusion that most PNe
originate from masses $\lesssim 2.5M_{\odot}$.
Unlike C, O, and Ne, the abundances of P, S, and Fe are affected
by neutron-capture reactions in the He-intershell.
Depending on assumptions made about
the efficiency of the $^{13}$C($\alpha$,n)$^{16}$O neutron source
\citep[see, e.g.,][]{busso01,herwig05}, P is predicted to be
produced by factors of $\sim 1.5$ to 25 times the solar value,
whereas the intershell S abundances are 0.6 to 0.9 times solar
depending on the AGB model \citep{werner06,karakas09}.

To account for the discrepancy between the observed PG 1159
abundances and the predicted intershell composition,
\citet{herwig00} adopted AGB models with overshoot into the
C-O core, that has a composition which is roughly 50\% carbon
and 50\% oxygen.  The inclusion of diffusive convective
overshoot at the inner edge of the flash-driven convective
pocket during a TP dredges up some of the C-O material and
leads to large increases in the C and O abundances in the
intershell \citep{herwig00}, in line with the PG 1159 observations.
The amount of overshoot applied is still however an uncertain
free parameter \citep{herwig00}.

\begin{center}
\begin{figure}[h]
\includegraphics[width=11cm,angle=270]{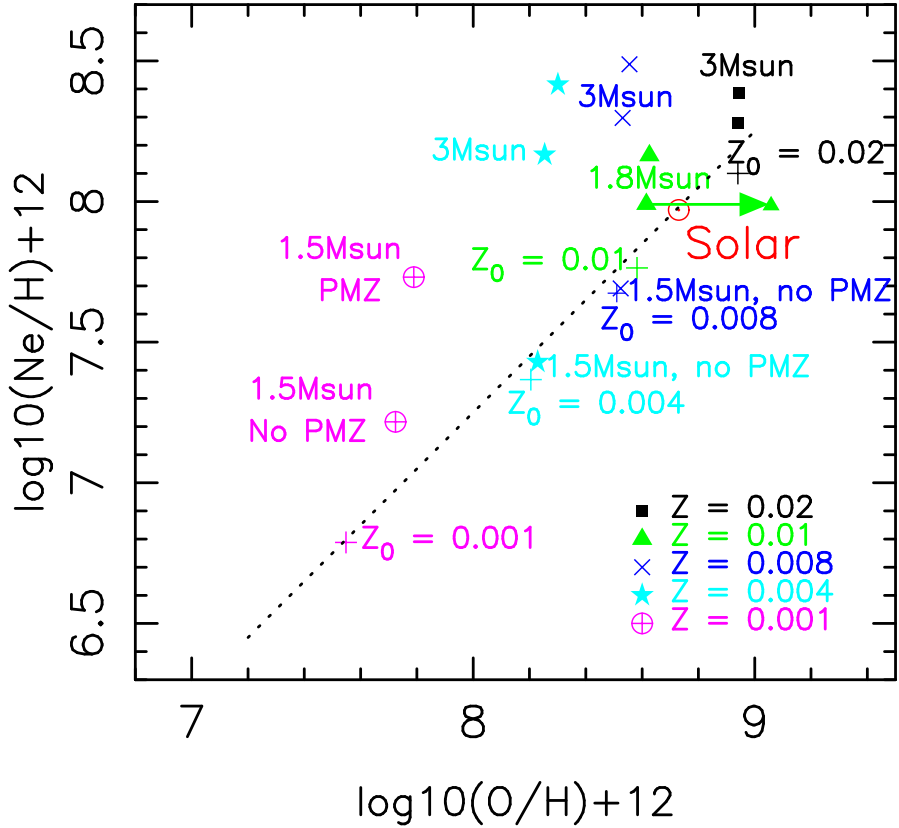}
\caption{The predicted $\log \epsilon$(O/H) versus 
$\log \epsilon$(Ne/H) abundances from a selection of
AGB models. The composition of the model is sampled from
the surface of the star at the very tip of the AGB. The solar composition 
is shown, and the approximate relationship between O and Ne abundances 
from HII regions (dotted line).  Initial abundances used in the model
calculations are shown by the cross symbols 
for each respective metallicity. The legend shows which symbols
correspond to models of a given metallicity. Predictions include: 3$M_{\odot}$, 
$Z=0.02$ models with and without a partially mixed zone (solid 
black squares); a 1.8$M_{\odot}$, $Z=0.01$ model with and without a 
partially mixed zone (solid green triangles); 
1.5$M_{\odot}$ and 3$M_{\odot}$ models of $Z=0.008$ (blue diagonal
crosses); 1.5$M_{\odot}$ and 3$M_{\odot}$ models of $Z=0.004$ (solid
aqua stars); and a 1.5$M_{\odot}$, $Z=0.001$ model with 
and without a partially mixed zone (magenta circle with cross-hatch).
In each case the upper symbol for the same mass corresponds to
the model with a partially mixed zone, as these have the highest
Ne abundances. We also include results of the synthetic AGB model with an 
O intershell abundance of 20\% (by mass), connected by 
the arrow to the model with a standard intershell 
composition (0.8\%  by mass), see the text for details.}
\label{neon}
\end{figure}
\end{center}

\begin{center}
\begin{figure}[h]
\includegraphics[width=11cm,angle=270]{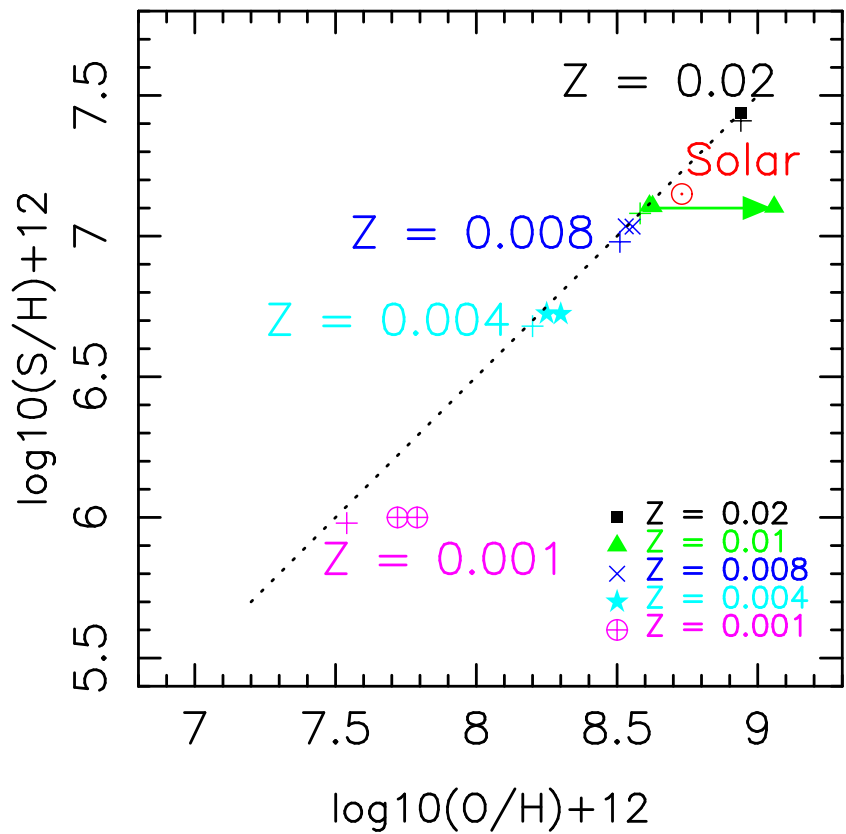}
\caption{Same as Fig.~\ref{neon} but for the 
predicted $\log \epsilon$(O/H) versus 
$\log \epsilon$(S/H). Here there is no clear distinction 
between models of different mass or for models 
with and without a partially mixed zone, except at $Z = 0.001$
(where we only include results at 1.5$M_{\odot}$).
Also included are results of the synthetic AGB model 
with an O intershell abundance of 20\% (by mass), connected
by the arrow to the model with a standard intershell 
composition (0.8\%  by mass); see text for details.}
\label{sulfur}
\end{figure}
\end{center}

Figs.~\ref{neon} and~\ref{sulfur} show the predicted abundances
of oxygen, neon, and sulfur from the AGB models
[the $Z=0.02, 0.008, 0.004$ models from \citet{karakas10a};
the $Z=0.01$ model from \citet{karakas10b}; and the
$Z=0.001$ model is from \citet{alves11}].
The surface abundances of the AGB models are taken at the tip of
the AGB, after the last computed thermal pulse and are taken
to be representative of the composition of the envelope when it
is ejected from the star.  In each figure we show
12+log(O/H) abundances on the $x$-axis versus 12+log(Ne/H)
(Fig.~\ref{neon}) or 12+log(S/H) (Fig.~\ref{sulfur}).

In Fig.~\ref{neon} we see that there is considerable spread
in the Ne abundances predicted from the AGB models.  The
models that produce the most Ne are those partial mixing zones
and those models with masses of $3M_{\odot}$, which have efficient
TDU \citep{karakas03a}. Studies of the $s$-process in low-mass
AGB stars suggest that protons need to be partially mixed into the
top layers of the He-intershell in order to efficiently activate
 the main neutron producing reaction
$^{13}$C($\alpha,n$)$^{16}$O \citep[e.g.,][]{gallino98}.
In the AGB models shown in Fig.~\ref{neon} and described
for the most part in \citet{karakas10a}, an
exponentially decreasing proton profile is introduced
into the top $\sim$10-15\% of the He-intershell
(a partially mixed zone or PMZ). The protons
are captured by the abundant $^{12}C$ to form a $^{13}$C pocket
(and usually a $^{14}$N pocket), with neutrons released by the
$^{13}$C($\alpha,n$)$^{16}$O reaction. The extra Ne
production arises from the extra $^{14}$N, which is converted
to $^{22}$Ne during convective thermal pulses.
From Fig~\ref{neon} we can see that the effect of this
is small at higher metallicities or in lower masses
(e.g.,  the 1.8$M_{\odot}$, $Z=0.01$).  The observational
data for O and Ne (see Fig.~\ref{nevo_initial}) shows some spread, and that
most of the lower mass models ($< 3M_{\odot}$) fall within
that range. Initial mass function considerations, which
favor lower mass stars, together with these predictions
suggest that the majority Galactic disk PNe evolved from stars
of $\lesssim 2M_{\odot}$. There are a few PNe
with higher Ne abundances and these may have evolved from
higher mass progenitors of $\sim 3 M_{\odot}$. At the lowest
metallicities shown in Fig.~\ref{neon} we similarly conclude
that most PNe with $\log$(O/H)$ < 8$ likely have evolved from
progenitors less massive than 1.5$M_{\odot}$ and/or did not
experience efficient third dredge-up.

In Fig.\ref{sulfur} we show the predicted abundances
of O and S from the same set of AGB models shown in
Fig.~\ref{neon}. In contrast to the predicted spread in Ne,
there is no predicted spread in sulfur present in any of
the models.  We can conclude that AGB models do not result
in a net production (or destruction) of S. This is consistent
with other AGB calculations
\citep[e.g.,][]{forestini97,cristallo09} and suggests that
current AGB models cannot solve the S anomaly problem.

However, the lowest metallicity models show an increase of O and
move to the right in the O versus S plot. What if all AGB models,
regardless of metallicity, dredged-up sufficient O to move
them to the right on the O versus S plot? This could be achieved
if convective overshoot from the C-O core increased the O content
of the He-intershell as discussed above for PG 1159 stars.
In \citet{karakas10b} the synthetic evolution of a 1.8$M_{\odot}$,
$Z = 0.01$ model with higher C and O intershell abundances
was computed (it was assumed  that the neon and sulfur abundance are
not changed by this overshoot).
In Figs.~\ref{neon} and~\ref{sulfur} we
show the results of the synthetic 1.8$M_{\odot}$, $Z=0.01$
AGB model with an O intershell abundance of 20\% (by mass),
compared to the model with a standard intershell composition
(the arrow connects the models and shows the direction caused
by the shift in O).  This model moves to the right of the
line in both figures, and could provide an explanation
for the S anomaly.

Is there a need for such overshoot in AGB models?
As indicated previously, plots of sulfur versus neon and 
sulfur versus argon for PNe compared to HII regions show an 
offset in the sulfur abundance by the same amount compared to 
the plot of sulfur versus oxygen. This indicates
that the problem lies with the sulfur abundances and not
with the AGB nucleosynthesis models. However, there 
are arguments for the inclusion of convective overshoot
from the C-O core into the envelope. These are 
largely based on the composition of the post-AGB
PG 1159 stars. These objects are He-rich, likely caused by 
a late or very late thermal pulse \citep[e.g.,][]{bloecker01}.
Another supporting piece of evidence is the apparent correlation
between oxygen isotope ratios in AGB stars \citep{harris85b,harris87}.
Not only are they correlated with each other but appear to
be correlated with the observed carbon and $s$-process element 
abundances \citep{harris87}. Given  the large
observational uncertainties in the derived oxygen isotope ratios
it is difficult to draw firm conclusions \citep[see also][]{smith90a}.
Lastly, the high $^{12}$C intershell abundances that would 
result from overshoot into the C-O core changes the
neutron exposure and the resulting s-process abundance
distribution such that they are no longer consistent with
mainstream SiC grains measurements \citep{lugaro03a}.
This suggests that such overshoot is not common in AGB
stars that become C and $s$-process rich.

In summary, current AGB models are unable to account for the
S anomaly observed in PNe. The inclusion of overshoot from the
C-O core into the He-intershell could provide a tentative 
explanation but only when considering the correlation between
O and S. Overshoot does not help clarify the offset 
present between S and Ne and S and Ar. Furthermore, there are 
concerns as to the extent of convective overshoot into
the C-O during the AGB phase.

\section{Summary and Conclusions}\label{summary}

The sulfur anomaly is the tendency of planetary nebula sulfur abundances determined from collisionally excited optical line measurements of S$^+$ and S$^{+2}$ and an ionization correction factor to fall significantly below expected levels. In order to understand its origin, we have quantified the sulfur anomaly by computing a sulfur deficit, i.e., the difference between the observed and expected S abundances, and attempted to relate this value to a broad assortment of other factors relevant to PNe. These include directly measurable properties of the central star and nebula, problems related to determining the contribution of unobservable ionization stages of S to the total elemental abundance (including the effect of inappropriate rates of dielectronic recombination in nebular models used to determine the ionization correction factor for S), sequestration of S into dust and/or molecules, and nucleosynthesis in AGB stars which might either produce S-poor or O-rich material.

Nearly all of these factors can be eliminated from consideration as offering a solution to the problem. We find no evidence that the gap between observed and expected S abundances in PNe is statistically related to any of the 24 nebular or central star properties which we considered with the one exception of $\log(S^{+3}/S^{+2})$ which appears to exhibits a positive correlation. A careful comparison of sulfur deficits and possible IR signatures fails to reveal any indication of sequestration of sulfur into dust or molecules. Likewise for nucleosynthesis scenarios. Using AGB evolution models, we considered in detail if sulfur deficits could be related to either sulfur destruction or oxygen production during the AGB stage of stellar evolution. Here again, we obtained a null result.

The remaining factors which we considered are associated with the sulfur ionization correction factor and the ability to correct for abundances of sulfur ions higher than S$^{+2}$. We presented photoionization models of three PNe representing a range in sulfur deficit. The models appeared to be consistent with true gas-phase abundance shortfalls, leaving us with the option of adopting that as the solution or speculating that the models (from which the ICFs are formulated) are failing to account for significant occupation of higher ionization stages. The latter seemed less radical, especially since our consideration of dust/molecules and AGB nucleosynthesis strongly argues against a truly reduced S abundance. Therefore, we further pursued the ICF angle. 

We evaluated the impact of under- or overestimated dielectronic recombination rates in altering the sulfur ion population distribution and thereby producing a spurious model-based ICF. We did this by computing large grids of photoionization models in which we tested factor-of-two changes in the rates and compared output line strength ratios with their observed counterparts. We concluded that no reasonable amount of uncertainty in DR rates could explain the sulfur deficits.

Finally, we looked carefully at the PN sulfur abundances derived from direct observation of S$^{+3}$ at 10.5$\mu$m. Having this measurement presumably reduces or eliminates the need to employ an ICF. We were unable to show with statistical certainty that the sulfur deficit for each object correlates with its S$^{+3}$ abundance. On the other hand, a visual comparison of positions in the $\epsilon(S)$ versus $\epsilon(O)$ plane of objects with S abundances inferred from the ICF method and those whose abundances are derived with the IR method show that objects in the latter category generally have smaller sulfur deficits. In fact once the S$^{+3}$ abundances are accounted for, the average size of the remaining deficits for the IR objects resemble the expected size of the abundance scatter, although the deficits themselves are not completely eliminated. It would appear from this part of the analysis, then, that the sulfur deficit is related to the failure of ICFs commonly in use to correct for the true amount of S$^{+3}$ (and possibly higher ionization stages) present in these objects. This notion is supported by the direct correlation we found in \S\ 2 between $\log(S^{+3}/S^{+2})$ and the size of the sulfur deficit. Statistical confirmation of this should follow once additional IR observations are performed. Independent theoretical verification of this point could come from model computations such as those in progress by \citet{jacob11}. By introducing a negative density gradient into his CLOUDY models, Jacob boosts the relative amounts of the higher ionization stages of sulfur. His approach stands in contrast to the usual use of a constant gas density over the entire nebula, i.e., the common assumption made when using models to infer the form of the sulfur ICF.

\acknowledgments

RBCH gratefully acknowledges support from NSF (AST-0806490). AS is grateful for a Big XII Fellowship that enabled her to visit the University of Oklahoma. AIK is grateful for the support of the NCI National Facility at the ANU. GJF acknowledges support by NSF (0908877), NASA (07-ATFP07-0124, 10-ATP10-0053, and 10-ADAP10-0073) and STScI (HST-AR-12125.01 and HST-GO-12309). We also thank Karen Kwitter, Bruce Balick, Rolf Jacob, Ed Baron, Harriet Dinerstein, and Joe Rodgers for useful discussions and Matthew Husek, Aven King, and Alice Sady for their help in computing Cloudy models. This study made use of the PN abundance database compiled by Dr. Karen Kwitter and her students at Williams College, Vivienne Baldassare, Connor Dempsey, and Brian Kirk, as well as the NIST Atomic Spectra Database. Finally, we thank the referee for a careful reading of the manuscript and for making substantive suggestions which helped to improve the paper.

\appendix
\section{Correlation Coefficients}
When looking for correlations amongst various parameters it is important to 
define what constitutes a correlation. The strength of a correlation can
be measured via the correlation coefficient ($r$). 
Often it is 
assumed that any linear regression fit for which r$>0.5$ is an acceptable 
criterion for a significant correlation \citep[see e.g.,][]{speck97,dijkstra05}. 
At the same time \citet{thompson06} and \citet{guhaniyogi11}
define a significant correlation as a set of points for which 
$r > 0.7$ \citep[see also][]{dijkstra05}.

The `{\em sample correlation coefficient}', 
$r$, measures both the direction and strength of the relationship 
between an independent variable $x$ and a dependent variable $y$.  
%
The value of the sample correlation coefficent is $-1 < r < 1$ such that if 
$ r \sim 0$
there is no relationship between $x$ and $y$. The sign ($+$ or $-$) of $r$ 
merely designates whether the relationship between $x$ and $y$ is positive or 
negative (i.e. it defines the direction of the slope of the trend.). 
%
According to \citet{rumsey03}, most statisticians like to see $|r| > 0.6 $ in 
order to consider a correlation significant, while \citet{thompson06} stated 
$r$ should be $> 0.7$, in order to be considered a correlation. For the analysis here we choose the former, $|r| > 0.6 $.


\pagebreak

\begin{deluxetable}{lccc}
\tablecolumns{4}
\tablewidth{0pc}
\tabletypesize{\scriptsize}
\tablenum{1}
\tablecaption{Sulfur Deficit Correlation Tests}
\tablehead{
\colhead{Independent Variable\tablenotemark{1}} &\colhead{No. of Objects} &\colhead{r} &
\colhead{P}}

\startdata

c & 163 & -0.35 & 3.3E-6 \\
log(He$^{+2}$/He$^+$) & 140 & +0.15 & 0.075 \\
ICFO & 161 & +0.19 & 0.015 \\
ICFS & 163 & -0.024 & 0.76 \\
log Diameter & 101 & +0.28 & 0.004 \\
CPN Mass & 95 & -0.22 & 0.030 \\
log L$_\ast$ & 106 &+0.11 & 0.26 \\
Ne Deficit & 151 & -0.13 & 0.11 \\
12 + log(O/H) & 163 & +0.076 & 0.33 \\
log(O$^{+2}$/O$^+$) & 161 & +0.047 & 0.55 \\
log(S$^{+3}$/S$^+2$) & 25 & +0.60 & 1.43E-3 \\
log(Ar$^{+4}$/Ar$^+3$) & 73 & -0.067 & 0.57 \\
log(Ar$^{+3}$/Ar$^+2$) & 27 & +0.29 & 0.14 \\
log(O$_{obs}$-O$_{grad}$) & 124 & +0.38 & 9.1E-6 \\
R$_g$(kpc) & 124 & +0.21 & 0.018 \\
T(O III) & 151 & +0.058 & 0.48 \\
T(O II) & 136 & +0.0051 & 0.95 \\
T(S II) & 89 & +0.053 & 0.62 \\
T( S III) & 147 & +0.33 & 3.44E-5 \\
T(N II) & 145 & -0.11 & 0.19 \\
N$_e$(S II) & 147 & -0.088 & 0.29\\
N$_e$(Cl III) & 65 & +0.076 & 0.54\\
Excitation Class ($\lambda$5007) & 161 &+0.11 & 0.16\\
Excitation Class ($\lambda$4686) & 139 & +0.07 & 0.44

\enddata
\tablenotetext{1}{Variables in order from top to bottom are: reddening factor, log of the ion ratio, oxygen ionization correction factor, sulfur ionization correction factor, log of the nebular diameter, central star mass, log of the central star's luminosity, neon deficit (defined in the same way as the sulfur deficit), log oxygen abundance, log of the ion ratio (4 rows), log of the difference between a PN's measured O abundance and the ambient interstellar abundance at its location along the disk, PN's galactocentric distance along the disk, electron temperatures derived from the lines indicated (5 rows), electron densities derived from the lines indicated (2 rows), and excitation class based upon the emission line indicated (2 rows).}
\label{correlations}
\end{deluxetable}

\begin{deluxetable}{lcccccc}
\tablecolumns{7}
\tablewidth{0pc}
\tabletypesize{\scriptsize}
\tablenum{2}
\tablecaption{Photoionization Models}
\tablehead{
\colhead{}&\multicolumn{3}{c}{Observed}&\multicolumn{3}{c}{Models}\\
\colhead{Object} &\colhead{O/H} &\colhead{S/H} &\colhead{SD}&\colhead{O/H}&\colhead{S/H}&\colhead{SD}}

\startdata
IC 4593 & 8.62 & 6.66 & 0.42 & 8.65 & 6.55 & 0.56 \\
Hu 2-1 & 8.43 & 6.20 & 0.68 & 8.39 & 6.16 & 0.68 \\
NGC 3242 & 8.57  & 7.00 & 0.02  &  8.61 & 6.85 & 0.22
\enddata
\label{models}
\end{deluxetable}

\begin{deluxetable}{lcccccccc}
\tablecolumns{9}
\tablewidth{0pc}
\tabletypesize{\scriptsize}
\tablenum{3}
\tablecaption{Comparison of ICF and IR Results}
\tablehead{
\colhead{Object} &\colhead{$\epsilon(O)$\tablenotemark{1}} &\colhead{$\epsilon(S_{exp}$)\tablenotemark{1}} &\colhead{$\epsilon(S_{icf}$)\tablenotemark{1}}&\colhead{SD$_{icf}$}&\colhead{$\epsilon(S_{ir}$)\tablenotemark{1}}&\colhead{SD$_{ir}$}&\colhead{$\epsilon(S_{ir})-\epsilon(S_{icf}$)}&\colhead{S$^i/S_{exp}$\tablenotemark{2}}
}
\startdata
\cutinhead{SMC\tablenotemark{3}}
SMP8	&	7.88	&	6.33	&	5.97	&	0.36	&	6.11	&	0.22	&	0.14	&	0.40	\\
SMP11	&	8.02	&	6.47	&	6.38	&	0.09	&	6.28	&	0.19	&	-0.10	&	0.35	\\
SMP13	&	8.06	&	6.51	&	5.92	&	0.59	&	5.96	&	0.55	&	0.04	&	0.72	\\
SMP17	&	8.21	&	6.66	&	6.07	&	0.60	&	6.15	&	0.51	&	0.08	&	0.69	\\
SMP24	&	8.06	&	6.51	&	6.11	&	0.40	&	6.11	&	0.40	&	0.00	&	0.60	\\
SMP27	&	8.00	&	6.45	&	5.96	&	0.49	&	5.84	&	0.61	&	-0.12	&	0.75	\\
\cutinhead{MWG\tablenotemark{4}}
N6210	&	8.68	&	7.14	&	6.85	&	0.29	&	6.87	&	0.27	&	0.02	&	0.46	\\
N3242	&	8.57	&	7.02	&	7.00	&	0.03	&	6.45	&	0.57	&	-0.55	&	0.73	\\
N6369	&	8.71	&	7.16	&	6.80	&	0.37	&	6.78	&	0.39	&	-0.02	&	0.59	\\
N2392	&	8.28	&	6.73	&	6.53	&	0.20	&	6.7	&	0.03	&	0.17	&	0.06	\\
IC2448	&	8.44	&	6.89	&	5.93	&	0.97	&	6.3	&	0.59	&	0.38	&	0.74	\\
M1-42	&	8.55	&	7.01	&	7.06	&	-0.05	&	7.45	&	-0.44	&	0.39	&	\nodata	\\
He 2-111	&	8.44	&	6.90	&	6.97	&	-0.08	&	7.18	&	-0.28	&	0.20	&	\nodata	\\
Hu 1-2	&	8.43	&	6.88	&	6.20	&	0.68	&	6.62	&	0.26	&	0.42	&	0.45	\\
IC418	&	8.28	&	6.73	&	6.92	&	-0.18	&	6.64	&	0.09	&	-0.27	&	0.18	\\
IC2165	&	8.40	&	6.86	&	6.48	&	0.38	&	6.65	&	0.21	&	0.18	&	0.38	\\
MZ3   	&	8.16	&	6.61	&	6.76	&	-0.16	&	7.00	&	-0.39	&	0.24	&	\nodata	\\
N2022	&	8.80	&	7.26	&	6.65	&	0.61	&	6.80	&	0.46	&	0.15	&	0.65	\\
N2440	&	8.62	&	7.07	&	6.46	&	0.61	&	6.67	&	0.40	&	0.21	&	0.60	\\
N5315	&	8.64	&	7.09	&	7.19	&	-0.10	&	7.08	&	0.01	&	-0.11	&	0.03	\\
N5882	&	8.68	&	7.14	&	6.96	&	0.18	&	7.11	&	0.03	&	0.15	&	0.06	\\
N6302	&	8.21	&	6.66	&	6.77	&	-0.11	&	6.89	&	-0.23	&	0.12	&	\nodata	\\
N6445	&	8.83	&	7.29	&	6.74	&	0.54	&	6.89	&	0.40	&	0.15	&	0.60	\\
N6537	&	8.32	&	6.77	&	6.98	&	-0.21	&	7.04	&	-0.27	&	0.06	&	\nodata	\\
N6741	&	8.79	&	7.24	&	6.77	&	0.47	&	7.04	&	0.20	&	0.27	&	0.37	\\
N6886	&	8.65	&	7.11	&	6.74	&	0.37	&	7.00	&	0.11	&	0.26	&	0.22	\\
N7027	&	8.52	&	6.97	&	6.86	&	0.11	&	6.97	&	0.00	&	0.12	&	\nodata	\\
N7662	&	8.55	&	7.01	&	6.64	&	0.37	&	6.82	&	0.19	&	0.18	&	0.35	\\
DdDm-1	&	8.06	&	6.51	&	6.34	&	0.18	&	6.31	&	0.20	&	-0.03	&	0.37

\enddata
\tablenotetext{1}{$\epsilon(X)=12+log(X/H)$}
\tablenotetext{2}{$S^i/S_{exp}=1-\log^{-1}(-SD_{ir})$}
\tablenotetext{3}{\citet{shaw10}. Values for $\epsilon(S_{icf})$ and SD$_{icf}$ were derived from the S$^+$ and S$^{+2}$ ionic abundances listed in their paper and the sulfur ICF formulation of \citet{kwitter01}.}
\tablenotetext{4}{\citet{pbs10} and \citet[DdDm-1 only]{henry08}. For all but DdDm-1, oxygen abundances in column 2 were taken from the HKB04 survey, while the sulfur abundances were those determined by Pottasch \& Bernard-Salas.}
\label{ir}
\end{deluxetable}

%


%
%


\begin{thebibliography}{}



\bibitem[Ali et al. (1991)]{ali91}Ali, B., Blum, R. D., Bumgardner, T. E., Cranmer, S. R., Ferland, G. J., Haefner, R. I., \& Tiede, G. P. 1991, PASP, 103, 1182

\bibitem [Alves-Brito et al. (2011)]{alves11}Alves-Brito, A., Hau, George K. T., Forbes, D.A., Spitler, L.R., Strader, J., Brodie, J.P.., \& Rhode, K.L. 2011, \aap, in press

\bibitem[Asplund et al. (2009)]{asplund09}Asplund, M., Grevesse, N., Sauval, A.-J., \& Scott, P. 2009, \araa, 47, 481

\bibitem[Badnell (1991)]{badnell91}Badnell, N.R. 1991, \apj, 379, 356

\bibitem[Begemann et al. (1994)]{Begemann1994} Begemann, B. et al. 1994, \apjl, 423, L71

\bibitem[Bernard-Salas (2006)]{bs06}Bernard-Salas, J. 2006, in Planetary Nebulae in our Galaxy and Beyond, Proceedings of the International Astronomical Union, Symposium \#234. Edited by Michael J. Barlow and Roberto H. MŽndez. Cambridge: Cambridge University Press, pp.181-188

\bibitem[Bernard-Salas et al. (2008)]{jero}Bernard-Salas, J., Pottasch, S. R., Gutenkunst, S., Morris, P. W., Houck, J. R., 2008, \apj, 672, 274.

\bibitem[Bl{\"o}cker (2001)]{bloecker01}Bl{\"o}cker, T. 2001, \apss, 275, 1

\bibitem[Bl{\"o}cker \& Schoenberner (1991)]{bloecker91}Bl{\"o}cker, T. \& Schoenberner, D. 1991, \aap, 244, L43

\bibitem[Boothroyd \& Sackmann (1988)]{boothroyd88d} Boothroyd, A.~I. \& Sackmann, I.-J. 1988, \apj, 328, 671

\bibitem[Bujarrabal et al. (1994)]{bujarrabal1994}Bujarrabal, V., Fuente, A., Omont, A., 1994, \aap, 285, 247.

\bibitem[Busso et~al. (2001)]{busso01}Busso, M., Gallino, R., Lambert, D.~L., Travaglio, C., \& Smith, V.~V. 2001, \apj, 557, 802

\bibitem[Busso et~al. (1999)]{busso99}Busso, M., Gallino, R., \& Wasserburg, G.~J. 1999, \araa, 37, 239

\bibitem[Cristallo et~al. (2009)]{cristallo09}Cristallo, S., Straniero, O., Gallino, R., Piersanti, L., Dom{\'{\i}}nguez, I., \& Lederer, M.~T. 2009, \apj, 696, 797

\bibitem[\protect\citeauthoryear{Dijkstra et al.}{2005}]{dijkstra05} Dijkstra C., Speck A.K., Reid R.B., Abraham P. 2005, \apj, 633, L133

\bibitem[Dopita \& Meatheringham (1990)]{dopita90} Dopita, M.A., \& Meatheringham, S.J. 1990, \apj, 357, 140

\bibitem[Dopita et~al. (1997)]{dopita97}Dopita, M.~A., Vassiliadis, E., Wood, P.~R., Meatheringham, S.~J., Harrington, J.~P., Bohlin, R.~C., Ford, H.~C., Stecher, T.~P., \& Maran, S.~P. 1997, \apj, 474, 188

\bibitem[Esteban et al. (2004)]{esteban04}Esteban, C., Peimbert, M., Garc{'i}a-Rojas, J., Ruiz, M.T., Peimbert, A., Rodr{'i}guez, M. 2004, \mnras, 355, 229

\bibitem[Ferland et al. (1998)]{ferland98}Ferland, G.J., Korista, K.T., Verner, D.A., Ferguson, J.W., Kingdon, J.B., \& Verner, E.M. 1998, \pasp, 110, 761

\bibitem[Forestini \& Charbonnel (1997)]{forestini97}Forestini, M. \& Charbonnel, C. 1997, \aaps, 123, 241

\bibitem[French (1981)]{french81}French, H.B. 1981, \apj, 246, 434

\bibitem[Gallino et~al. (1998)]{gallino98}Gallino, R., Arlandini, C., Busso, M., Lugaro, M., Travaglio, C.,
  Straniero, O., Chieffi, A., \& Limongi, M. 1998, \apj, 497, 388

\bibitem[Goebel \& Moseley (1985)]{Goebel1985} Goebel, J.H., Moseley, S.H., 1985, \apjl, 290, L35

\bibitem[Gon\c{c}alves et al. (2011)]{goncalves11}Gon\c{c}alves, D.R., Wesson, R., Morisset, C., Barlow, M., \& Ercolano, B. 2011, in Planetary Nebulae: An Eye to the Future, Proceedings IAU Symposium No. 283, A. Manchado \& L. Stanghellini, eds., Cambridge: Cambridge University Press, arXiv:1110.2709

\bibitem[Green et al. (2011)]{green11}Green, J.M., Braxton, K., Balick, B., \& Lutz, J. 2011, Bulletin of the American Astronomical Society, Vol. 43, 2011

\bibitem[\protect\citeauthoryear{Guha Niyogi et al.}{2011}]{guhaniyogi11} Guha Niyogi, S., Speck, A. K., Onaka, T., 2011, \apj, 733, 93.

\bibitem[Harris et~al. (1987)]{harris87}Harris, M.~J., Lambert, D.~L., Hinkle, K.~H., Gustafsson, B., \&
Eriksson, K. 1987, \apj, 316, 294

\bibitem[Harris et~al. (1985)]{harris85b}Harris, M.~J., Lambert, D.~L., \& Smith, V.~V. 1985, \apj, 299, 375

\bibitem[Henry et al. (2010)]{hk10}Henry, R.B.C., Kwitter, K.B., Jaskot, A.E., Balick, B., Morrison, M.A., \& Milingo, J.B. 2010, \apj, 724, 748

\bibitem[Henry et al. (2008)]{henry08}Henry, R.B.C., Kwitter, K.B., Dufour, R.J., \& Skinner, J.N. 2008, \apj, 680, 1162

\bibitem[Henry et al.\ (2004)]{hkb04} Henry, R.B.C., Kwitter, K.B., \& Balick, B. 2004, \aj, 127, 2284 (HKB04)

\bibitem[Herwig (2000)]{herwig00}Herwig, F. 2000, \aap, 360, 952

\bibitem[Herwig (2005)]{herwig05}Herwig, F. 2005, \araa, 43, 435

\bibitem[Hony, Waters, \& Tielens (2002)]{Hony2002} Hony, S., Waters, L.B.F.M., Tielens, A.G.G.M., 2002, \aap, 390, 533

\bibitem[Huggins \& Healy(1989)]{huggins1}Huggins, P.J., Healy, A.P. 1989, \apj, 346, 201.

\bibitem[Huggins et al.(1996)]{huggins2} Huggins, P.J., Bachiller, R., Cox, P., Forveille, T., 1996, \aap, 315, 284,

\bibitem[Izotov \& Thuan (1999)]{it99}Izotov, Y.I., \& Thuan, T.X. 1999, \apj, 511, 639

\bibitem [Jacob (2011)]{jacob11}Jacob, R. 2011, Poceedings of  IAU Symposium No. 283, Planetary Nebulae: An Eye to the Future, A. Manchado \& L. Stanghellini, eds., (Cambridge: Cambridge University Press), in press

\bibitem[Johnson \& Levitt (2006)]{johnson06} Johnson, M.D., \& Levitt, J.S. 2006, IAU Symposium Series, 234, 439

\bibitem[Karakas (2010)]{karakas10a}Karakas, A.~I. 2010, \mnras, 403, 1413

\bibitem[Karakas et~al. (2010)]{karakas10b}Karakas, A.~I., Campbell, S.~W., \& Stancliffe, R.~J. 2010, \apj, 713, 374

\bibitem[Karakas \& Lattanzio (2003)]{karakas03a}Karakas, A.~I. \& Lattanzio, J.~C. 2003, Publ. Astron. Soc. Aust., 20, 393

\bibitem[Karakas et~al. (2002)]{karakas02}Karakas, A.~I., Lattanzio, J.~C., \& Pols, O.R. 2002, Publ. Astron. Soc. Aust., 19, 515


\bibitem[Karakas et al. (2009)]{karakas09}Karakas, A.~I., van Raai, M.~A., Lugaro, M., Sterling, N.~C., \& Dinerstein, H.~L. 2009, \apj, 690, 1130

\bibitem[Kennicutt, Bresolin, \& Garnett (2003)]{kbg03}Kennicutt, R.C., Bresolin, F., \& Garnett, D.R. 2003, \apj, 591, 801

\bibitem[Kingsburgh \& Barlow (1994)]{kb94}Kingsburgh, R.L., \& Barlow, M.J. 1994, \mnras, 271, 257

\bibitem[Krugel(2008)]{krugel2008} Krugel, E.  2008, ``An Introduction to the Physics of Interstellar Dust'' Taylor \& Francis Group, LLC, New York, p387.

\bibitem[Kurucz (1991)]{kurucz91}Kurucz, R.L. 1991, in Proceedings of the Workshop on Precision Photometry: Astrophysics of the Galaxy, ed. A.C. Davis Philp, A.R. Upgren, \& K.A. James, (Schenectady: Davis), 27

\bibitem[Kwitter et al. (2012)]{kwitter12}Kwitter, K.B., Kehman, E.M.M., Balick, B., \& Henry, R.B.C. 2012, \apj, {\it in preparation}

\bibitem [Kwitter \& Henry (2011)]{kh11} Kwitter, K.B., \& Henry, R.B.C. 2011, in Planetary Nebulae: An Eye to the Future
Proceedings IAU Symposium No. 283, A. Manchado \& L. Stanghellini, eds., Cambridge: Cambridge University Press, arXiv:1109.2502

\bibitem[Kwitter \& Henry (2001)]{kwitter01} Kwitter, K.B., \& Henry, R.B.C. 2001, \apj, 562, 804

\bibitem[Leisy \& Dennefeld (2006)]{ld06}Leisy, P., \& Dennefeld, M. 2006, \aap, 456, 451L

\bibitem[Lodders \& Fegley(1995)]{lodders1995} Lodders, K., Fegley, B., 1995, Meteoritics, 30, 661

\bibitem[Lugaro et~al. (2003)]{lugaro03a}Lugaro, M., Herwig, F., Lattanzio, J.~C., Gallino, R., \& Straniero, O. 2003, \apj, 586, 1305


\bibitem[Matsuura et al.(2009)]{matsuura2009} Matsuura, M., Speck, A.K., Mchunu, B.M. et al. 2009, \apj, 700, 1067.

\bibitem[Milingo et al.\ (2010)]{mkhs} Milingo, J. B., Kwitter, K. B., Henry, R. B. C., \& Souza, S. P., 2010, \apj, 711, 619

\bibitem[Nuth et al.(1985)]{nuth1985} Nuth, J.A., et al. 1985, \apjl, 290, L41.

\bibitem[Omont et al.(1995)]{Omont1995} Omont, A. et al. 1995, \apj, 454, 819

\bibitem[Omont et al.(1993)]{omont1993} Omont, A., Lucas, R., Morris, M., Guilloteau, S., 1993, \aap, 267, 490.

\bibitem[Osterbrock \& Ferland (2006)]{osterbrock06}Osterbrock, D.E., \& Ferland, G.J. 2006, Astrophysics of Gaseous Nebulae and Active Galactic Nuclei--2nd edition, University Science Books 


\bibitem[Peimbert \& Costero (1969)]{pc69}Peimbert, M., \& Costero, R. 1969, Bol. Obs. Tonantzintla Tacubaya, 5, 3

\bibitem[Pottasch \& Bernard-Salas (2006)]{pbs06}Pottasch, S.R., \& Bernard-Salas, J. 2006, \aap, 457, 189

\bibitem[Pottasch et al. (2007)]{pottasch07}Pottasch, S.R., Bernard-Salas, J., \& Roellig, T.L.  2007, \aap, 471, 865

\bibitem[Pottasch et al. (2008)]{pottasch08}Pottasch, S.R., Bernard-Salas, J., \& Roellig, T.L. 2008, \aap, 481, 393

\bibitem[Pottasch \& Bernard-Salas (2010)]{pbs10}Pottasch, S.R., \& Bernard-Salas, J. 2010, \aap, 517, A95

\bibitem[Press et al. (2003)]{press03} Press, W.H., Teukolsky, S.A., Vetterling, W.T., \& Flannery, B.P. 2003, Numerical Recipes in Fortran 77, (Cambridge: Cambridge University Press), pg.~632

\bibitem[Rauch (1997)]{rauch97}Rauch, T. 1997, \aap, 320, 237

\bibitem[Rodr{\'i}guez \& Delgado-Inglada (2011)]{rodriguez11}Rodr{\'i}guez, M., \& Delgado-Inglada, G. 2011 in IAU Symposium 283, Planetary Nebulae: An Eye to the Future, A. Manchado \& L. Stanghellini, eds., (Cambridge: CUP), in press, astro-ph/1109.1861

\bibitem[\protect\citeauthoryear{Rumsey}{2003}]{rumsey03} Rumsey, D. 2003, ``Statistics for Dummies'' (Indianapolis: Wiley)

\bibitem[Shaver et al. (1983)]{shaver83}Shaver, P.A., McGee, R.X., Newton, L.M., Danks, A.C., \& Pottasch, S.R. 1983, \mnras, 204, 53

\bibitem[Shaw et al. (2010)]{shaw10}Shaw, R.A., Lee, T-H., Stanghellini, L., et al. 2010, \apj, 717, 562

\bibitem[Smith \& Lambert (1990)]{smith90a}Smith, V.~V. \& Lambert, D.~L. 1990, \apjs, 72, 387

\bibitem[\protect\citeauthoryear{Speck et al.}{1997}]{speck97} Speck A.K., Barlow M.J., Skinner, C.J., 1997, MNRAS, 288, 431.

\bibitem[Straniero et~al. (2003)]{straniero03}Straniero, O., Dom{\'{\i}}nguez, I., Cristallo, R., \& Gallino, R. 2003, Publ. Astron. Soc. Aust., 20, 389

\bibitem[Stasi{\'n}ska (1978)]{stasinska78}Stasi{\'n}ska, G. 1978, \aap, 66, 257

\bibitem[Thompson et al.(2006)]{thompson06}Thompson, G.D., Corman, A.B., Speck, A.K., Dijkstra,  2006, \apj, 652, 1654.

\bibitem[Wallerstein \& Knapp (1998)]{wallerstein98}Wallerstein, G. \& Knapp, G.~R. 1998, \araa, 36, 369

\bibitem[Werner \& Herwig (2006)]{werner06}Werner, K. \& Herwig, F. 2006, \pasp, 118, 183

\bibitem[Werner \& Rauch (1994)]{werner94}Werner, K. \& Rauch, T. 1994, \aap, 284, L5

\bibitem[Werner et~al. (2005)]{werner05}Werner, K., Rauch, T., \& Kruk, J.~W. 2005, \aap, 433, 641

\bibitem[Werner et~al. (2009)]{werner09}Werner, K., Rauch, T., Reiff, E., \& Kruk, J.~W. 2009, \apss, 320, 159

\bibitem[Whittet (1992)]{whittet1992}Whittet, D. C. B., 1992, ``Dust In The Galactic Environment'', IoP Publishing.










\end{thebibliography}
\end{document}